\documentclass[aps,prb,twocolumn,amsmath,amssymb,superscriptaddress,10pt]{revtex4-2}

\usepackage[pdftex]{graphics}
\usepackage{graphicx}
\usepackage{xcolor}
\usepackage{physics}
\usepackage{ulem}

\begin{document}

\title{Between Mott and cluster Mott: spin-orbit entangled dimer singlets in Ba$_3$CeRu$_2$O$_9$}

\author{L. P\"atzold}
\affiliation{Institute of Physics II, University of Cologne, 50937 Cologne, Germany}
\author{A. Sandberg}
\affiliation{Department of Physics, Stockholm University, AlbaNova University Center, SE-106 91 Stockholm, Sweden}
\author{H. Schilling}
\affiliation{\mbox{Sect.\ Crystallography, Institute of Geology and Mineralogy, University of Cologne, 50674 Cologne, Germany}}
\author{H. Gretarsson}
\affiliation{PETRA III, Deutsches Elektronen-Synchrotron DESY, 22607 Hamburg, Germany}
\author{E. Bergamasco}
\author{M. Magnaterra}
\affiliation{Institute of Physics II, University of Cologne, 50937 Cologne, Germany}
\author{P.~Becker}
\affiliation{\mbox{Sect.\ Crystallography, Institute of Geology and Mineralogy, University of Cologne, 50674 Cologne, Germany}}
\author{P. H. M. van Loosdrecht}
\affiliation{Institute of Physics II, University of Cologne, 50937 Cologne, Germany}
\author{J. van den Brink}
\affiliation{Institute for Theoretical Solid State Physics, IFW Dresden, 01069 Dresden, Germany}
\affiliation{Institute for Theoretical Physics and W\"urzburg-Dresden Cluster of Excellence ct.qmat, Technische Universit\"at Dresden, 01069 Dresden, Germany}
\author{M.~Hermanns}
\affiliation{Department of Physics, Stockholm University, AlbaNova University Center, SE-106 91 Stockholm, Sweden}
\affiliation{Stockholm University, SE-106 91 Stockholm, Sweden}
\author{M. Gr\"{u}ninger}
\affiliation{Institute of Physics II, University of Cologne, 50937 Cologne, Germany}

\begin{abstract}
The hexagonal $4d$ ruthenates Ba$_3$$M$Ru$_2$O$_9$ host structural dimers and exhibit a
delicate balance of competing interactions. Hund's coupling, trigonal crystal-field splitting, 
and hopping for $a_{1g}$ and $e_g^\pi$ orbitals all fall within a narrow energy window. 
This yields a series of possible ground states, 
ranging from the localized Mott limit with (anti-)ferromagnetic exchange coupling 
via orbital-selective behavior to the cluster Mott limit with quasimolecular orbitals that 
are delocalized over the two dimer sites. 
Using resonant inelastic x-ray scattering, we show that Ba$_3$CeRu$_2$O$_9$ with 
four holes per dimer resides in the intricate crossover regime between the localized Mott case 
and the quasimolecular limit. 
The spin-orbit entangled singlet ground state predominantly shows a Mott-like charge distribution 
with two holes per Ru site. 
At the same time, spin and orbital occupation contradict an exchange-based Mott scenario 
but agree with a cluster Mott approach. A quasimolecular trial wave function describes more than 
70\,\% of the ground state. In this crossover regime, small changes of, e.g., the crystal field 
may strongly affect the character of electronic states.
\end{abstract}

\date{April 8, 2026}

\maketitle

\noindent
\textbf{INTRODUCTION}
\\
In correlated transition-metal compounds, the entanglement of spins and orbitals 
and their interplay with other degrees of freedom give rise to an intriguing variety 
of properties and phases \cite{Khomskiibook,Streltsov17,Khomskii21,Streltsov16,Streltsov14}. 
The cornucopia of different crystal structures and substitutions offers the opportunity 
to realize  different parameter regimes and to tune the material properties.
A prominent example are compounds with $4d^4$ Ru$^{4+}$ ions. 
The rich phase diagram of layered Sr$_{2-x}$Ca$_x$RuO$_4$ includes 
the highly controversial unconventional superconductivity in 
Sr$_2$RuO$_4$ \cite{Ishida98,Pustogow19,Maeno24}
and a temperature-driven 
metal-insulator transition and antiferromagnetic order in 
Ca$_2$RuO$_4$ \cite{Nakatsuji97,Braden98}. 
For intermediate $x$, an orbital-selective Mott transition has been discussed  \cite{Anisimov02,Streltsov17,Neupane09}, where the degree of Mott localization depends on 
the orbital character.
For well separated Ru ions as in cubic K$_2$RuCl$_6$, spin-orbit coupling $\zeta$ yields a 
nonmagnetic $J$\,=\,0 ground state \cite{Takahashi21}. 
The competition of $\zeta$ and tetragonal crystal-field splitting $\Delta_{\rm tet}$ 
has been discussed extensively in Ca$_2$RuO$_4$ \cite{Kunkemoeller15,Jain17,Zhang17,Gretarsson19,Sarte20,Mohapatra20,Feldmaier20,Vergara22}. 
From the perspective of large $\Delta_{\rm tet}$ lifting orbital degeneracy,
Ca$_2$RuO$_4$ can be viewed as an $S$\,=\,1 antiferromagnet 
in which $\zeta$ causes a large single-ion anisotropy \cite{Kunkemoeller15,Zhang17}. 
The alternative scenario of excitonic magnetism \cite{Khaliullin13}
starts from large $\zeta$ and local $J$\,=\,0 moments and considers condensation 
of a dispersive, magnetic excited state. 
In this case, one expects a longitudinal magnon that has been discussed as being 
equivalent to a Higgs mode \cite{Jain17}. In fact, the local $4d^4$ ground 
state is a $J$\,=\,0 singlet for any $\Delta_{\rm tet}/\zeta$, and sizable 
$\Delta_{\rm tet}$ facilitates condensation in this picture.

\begin{figure}[t]
	\centering
	\includegraphics[width=0.95\columnwidth]{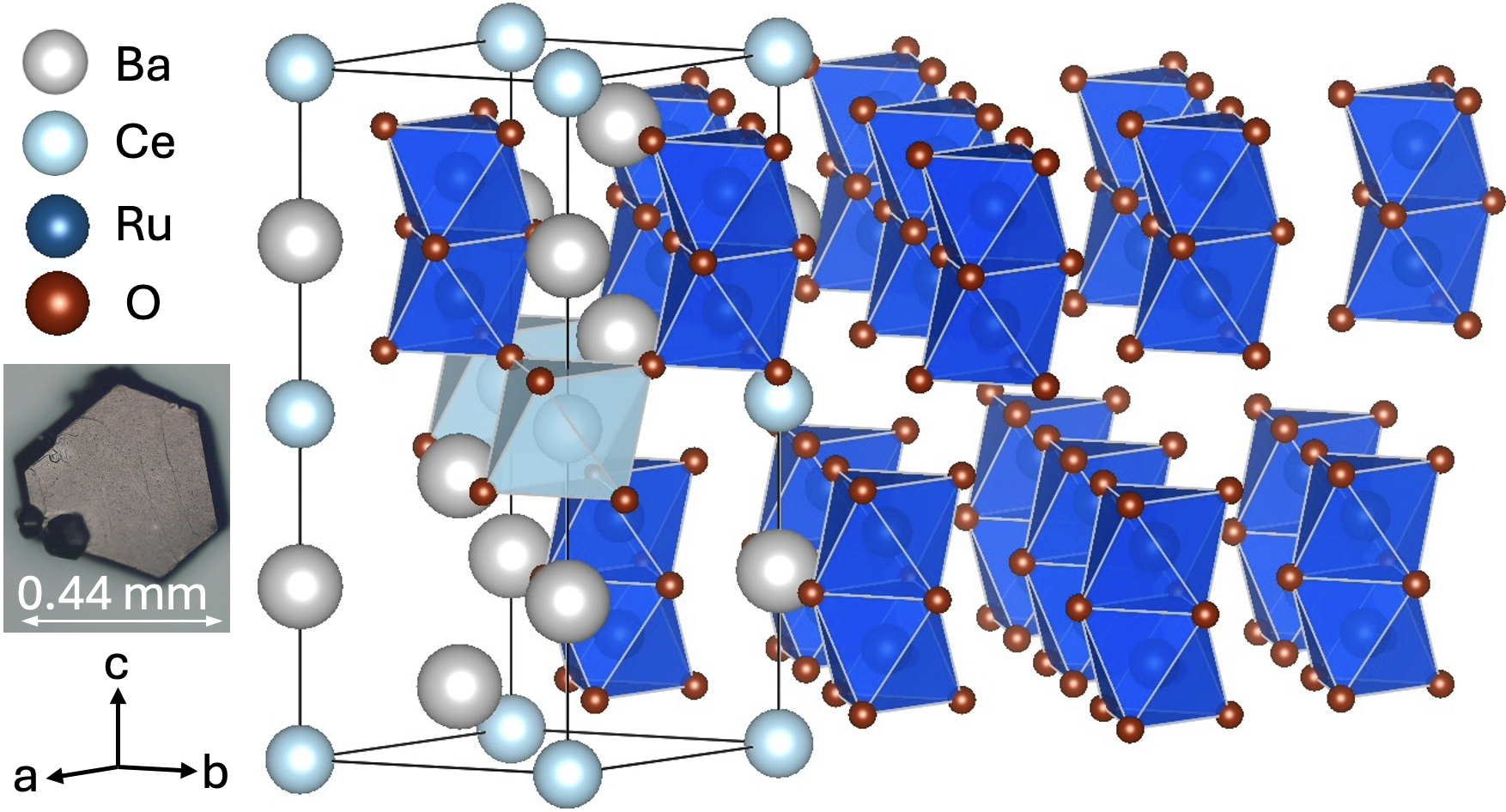}
	\caption{\textbf{Hexagonal crystal structure of Ba$_3$CeRu$_2$O$_9$.} 
    The unit cell hosts two distinct orientations of structural dimers, 
    each built by two face-sharing RuO$_{6}$ octahedra. The dimers grow along the $c$ axis 
    and form triangular layers. 
    Beyond the unit cell, only the Ru$_2$O$_9$ dimers are sketched for clarity.
    The photo shows one of the measured crystals.
} 
\label{fig:structure}
\end{figure}

Novel states of quantum matter may be realized in cluster Mott 
insulators \cite{Streltsov17,Khomskii21,Nikolaev21,Petersen23,Chen24,Jayakumar26}, 
which in essence are located in between Mott insulators and metals. 
In a cluster Mott insulator, Coulomb repulsion dominates over \textit{inter}-cluster hopping, 
causing an insulating state, while large intra-cluster hopping $t$ yields quasimolecular 
orbitals delocalized over a small cluster, e.g., a Ru dimer. 
The emergent internal degrees of freedom yield variable quasimolecular magnetic moments 
that can be tuned by electronic parameters \cite{Li20,Magnaterra24}.
In a simple cluster picture, one can distinguish the Mott limit, 
in which on-site Coulomb repulsion $U \gg t$ suppresses charge fluctuations 
between Ru sites, and the cluster Mott limit $t \gg U$. 
Such states may be realized in the large family of hexagonal perovskites 
with face-sharing RuO$_6$ octahedra \cite{Nguyen21}.
Compounds of $6H$-type Ba$_3$$M$Ru$_2$O$_9$ exist for many different $M$ ions,  
e.g., Na$^+$, Zn$^{2+}$, La$^{3+}$, and Ce$^{4+}$ \cite{Doi01,Doi02,Hinatsu03} 
and host structural Ru dimers, see Fig.\ \ref{fig:structure}.
The short intra-dimer Ru-Ru distance $d$\,$\approx$\,2.5 to 2.8\,\AA{} 
is expected  to yield large hopping \cite{Li20}. 
Concerning magnetism, the triangular layers of dimers show geometrical frustration 
in the case of antiferromagnetic couplings between dimers.
However, one first has to address the character of the possibly quasimolecular moments.  
In resonant inelastic x-ray scattering (RIXS) on the isostructural $5d$ iridates 
Ba$_3$$M$Ir$_2$O$_9$ ($M$\,=\,Ce, Ti, In) \cite{Revelli19,Revelli22,Magnaterra23Ti}, 
the quasimolecular character has been demonstrated, with 
the \mbox{(anti-)} bonding orbitals for large $\zeta$ being formed from 
spin-orbit entangled $j$ states. 
The spin-liquid candidate Ba$_3$InIr$_2$O$_9$ hosts quasimolecular $j$\,=\,3/2 
moments \cite{Revelli22} and shows persistent spin dynamics down to 20\,mK \cite{Dey17}.

The $4d$ ruthenates cover a different part of phase space, with smaller hopping, 
larger correlations, and smaller but still sizable spin-orbit coupling. 
For Ba$_3$$M$Ru$_2$O$_9$, 
an exact diagonalization (ED) study finds a variety of different states with anisotropic 
and temperature-dependent magnetic moments that depend on electron filling, correlations, 
and $\zeta$ \cite{Li20}. Experimentally, the reported behavior is diverse. 
For $M$\,=\,Na$^+$, charge order with a segregation into Ru$^{5+}$ and Ru$^{6+}$ dimers 
has been claimed \cite{Kimber12}, while a spin $S$\,=\,3/2 Mott insulator has been found 
for $M$\,=\,Zn$^{2+}$ with $4d^3$ Ru$^{5+}$ ions \cite{Hayashida25}. 
For $M$\,=\,La$^{3+}$, the results range from ferromagnetic double exchange interactions 
between the two Ru sites \cite{Senn13} via an orbital-selective $S$\,=\,3/2 
scenario \cite{Chen20} to a quasimolecular picture \cite{Yuan25}.
The electronic states are highly sensitive to small structural changes 
caused by different $M^{3+}$ ions \cite{Senn13,Ziat17,Chen20,Yuan25}. 
Finally, studies of polycrystalline $4d^4$ Ba$_3$CeRu$_2$O$_9$  find 
nonmagnetic behavior that has been discussed in the Mott limit \cite{Doi01} 
and in the quasimolecular limit \cite{Chen19}.

This diversity of partially conflicting results reflects 
the intertwined coupling of orbitals and spins on a dimer. Hopping $t$ 
does not only compete with on-site $U$ but also with Hund's coupling 
$J_{\rm H}$ and the trigonal crystal-field splitting $\Delta_{\rm trig}$. 
Furthermore, the trigonal symmetry splits the $t_{2g}$ manifold in $a_{1g}$ 
and $e_g^\pi$ orbitals with different hopping strengths $t_{a_{1g}}$ 
and $t_{e_g^\pi}$, promoting orbital-selective behavior 
\cite{Streltsov14,Streltsov16,Streltsov17,Khomskii21}. 
For $\zeta$\,=\,0, this yields a multitude of possible ground states which 
depend on the subtle hierarchy of electronic parameters \cite{Khomskii21}, 
and finite $\zeta$ further expands the picture \cite{Li20}.

Here, we address the electronic structure of the four-hole dimer compound 
Ba$_3$CeRu$_2$O$_9$ with RIXS at the Ru $L_3$ edge. 
We observe a rich excitation spectrum and a \textbf{q}-dependent modulation of 
the RIXS intensity. This allows us to determine the electronic parameters and 
the spin-orbit entangled singlet character of the ground state.
Using exact diagonalization, we characterize the different states that emerge for 
either small or large hopping and different crystal-field splittings. 
We show that Ba$_3$CeRu$_2$O$_9$ is best described as being located in the 
intriguing intermediate regime, combining aspects of the Mott limit and of 
the quasimolecular limit.

RIXS interferometry is a technique very well suited for revealing a possible 
cluster Mott character \cite{Revelli19,Revelli22,Magnaterra23Ti}.
In analogy to Young's double-slit experiment, the RIXS intensity of quasimolecular dimer 
excitations exhibits a sinusoidal interference pattern as a function of the transferred 
momentum \textbf{q}, arising from coherent scattering on the two dimer sites \cite{Revelli19}.
The interference pattern reveals the symmetry and character of the quasimolecular wavefunction, 
as demonstrated in the hard x-ray range for a series of $5d$ compounds with dimers, trimers, 
and tetrahedral clusters 
\cite{Revelli22,Magnaterra23Ti,Magnaterra24,Magnaterra25trimer,Katukuri22,Porter22,Kwon26}. 
For tender x-rays at the Ru $L_3$ edge, one has to cope with the smaller 
range of \textbf{q} that can be covered. 
However, RIXS interferometry has even been employed in the soft x-ray range, 
e.g., for O$_2$ molecules at the O $K$ edge \cite{Soederstroem23} 
and for magnetic excitations at the Fe $L$ edge \cite{Bhartiya25}. 
\\

\begin{figure}[t]
	\centering
	\includegraphics[width=0.94\columnwidth]{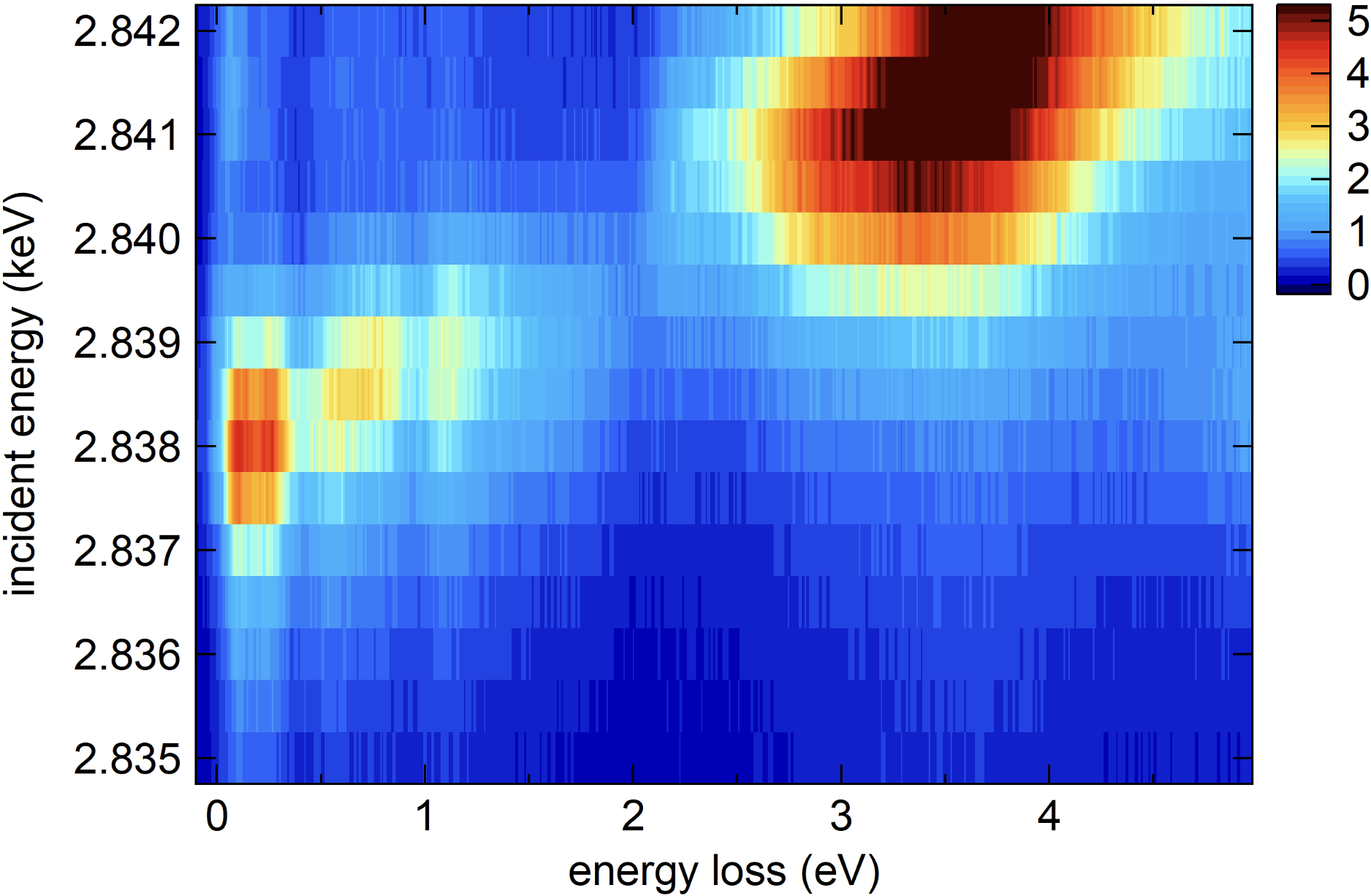}
	\caption{\textbf{Resonance map of Ba$_3$CeRu$_2$O$_9$ at 20\,K.\@} 
    The RIXS intensity is plotted for different incident energies across the Ru $L_3$ edge. 
    The data were taken on the (001) facet. Excitations from $t_{2g}$ to $e_g$ states are 
    peaking at about 3.5\,eV for $E_{\rm in}$\,$\approx$\,2.841\,keV, while intra-$t_{2g}$  
    excitations below 2\,eV energy loss are resonantly enhanced at 
    $E_{\rm in}$\,$\approx$\,2.838\,keV.\@ 
}
\label{fig:resonance}
\end{figure}

\begin{figure*}[t]
	\centering
 	\includegraphics[width=\textwidth]{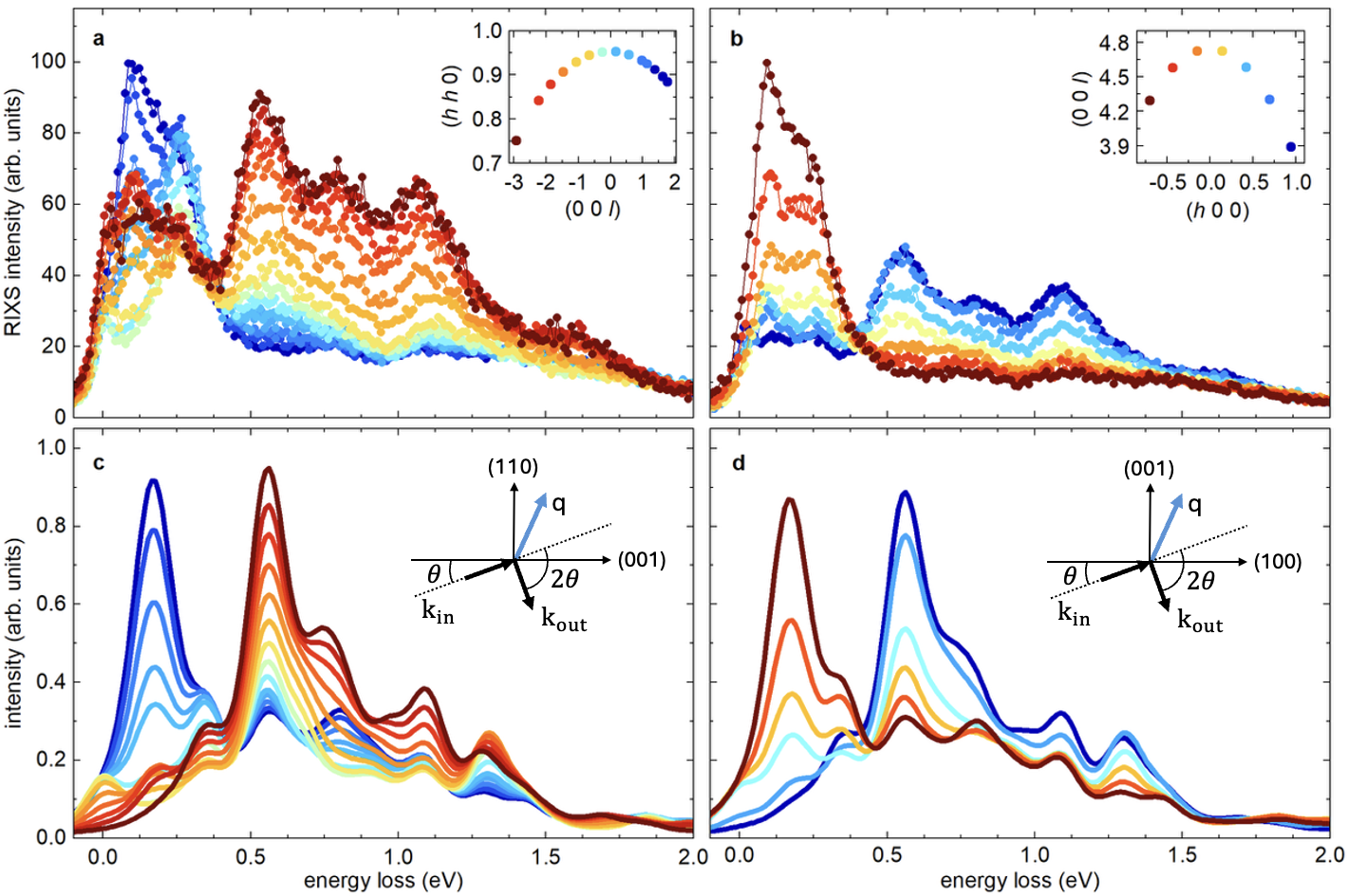}
	\caption{\textbf{RIXS spectra of Ba$_3$CeRu$_2$O$_9$.} 
    Data have been measured at 20\,K on \textbf{a} the (110) surface and 
    \textbf{b} the (001) surface for different angles of incidence $\theta$ 
    with fixed modulus $|\mathbf{q}|$, i.e., fixed scattering angle 
    $2\theta$\,=\,90$^\circ$, see sketches in \textbf{c} and \textbf{d}.
    The corresponding ($h$\,\,$k$\,\,$l$) values are given in the insets. 
    \textbf{c}, \textbf{d}: Calculated spectra for the orientations used in 
    \textbf{a}, \textbf{b}. 
    We employed $U$\,=\,2\,eV, $J_{\rm H}$\,=\,0.26\,eV, $\zeta$\,=\,0.15\,eV, 
    $\Delta_{\rm trig}$\,=\,0.27\,eV, $t_{a_{1g}}$\,=\,0.66\,eV, and $f$\,=\,$-0.45$ (i.e., $t_{e_{g}^\pi}$\,=\,$-0.30$\,eV). 
    For plotting, we further assumed a peak width of 90\,meV.\@  
}
\label{fig:spectra}
\end{figure*}

\noindent
\textbf{RESULTS}
\\
We studied Ru $L_3$-edge RIXS on single crystals of hexagonal Ba$_3$CeRu$_2$O$_9$, 
see Methods. We employed two different sample orientations, a (110) surface and 
a (001) surface. The resonance behavior of the RIXS intensity is presented in 
Fig.\ \ref{fig:resonance}. The spectra were measured on the (001) facet for 
incident energies between 2.835 and 2.842\,keV.\@ The most prominent RIXS feature 
is observed at about 3.5\,eV energy loss and corresponds to excitations from 
$t_{2g}$ to $e_g$ states. The excitation energy of 3.5\,eV provides an estimate 
of the cubic crystal-field splitting 10\,Dq. 
This $t_{2g}$-to-$e_g$ peak is resonantly enhanced at $E_{\rm in}$\,=\,2.841\,keV.\@ 
In the following, we focus on the intra-$t_{2g}$ excitations below 2\,eV energy loss 
that resonate at a lower energy of about 2.838\,keV.\@

RIXS spectra for the two different sample orientations are shown in 
Fig.\ \ref{fig:spectra}\textbf{a} and \textbf{b}. 
The data cover a broad range of the angle of incidence $\theta$ with 
fixed modulus $|\mathbf{q}|$, and the corresponding ($h$\,$k$\,$l$) values 
are depicted in the insets. The spectra are very rich with prominent RIXS peaks 
at about 0.10, 0.26, 0.53, 0.80, 1.1, and 1.6\,eV.\@  For all of them, the RIXS 
intensity strongly depends on $\theta$. As shown below, this originates from a 
\textbf{q} dependence of the intensity and from polarization effects.

In Ba$_3$CeRu$_2$O$_9$, a comprehensive description of the excitations 
requires to consider the interplay and competition of Coulomb interactions, 
hopping, crystal-field splitting, and spin-orbit coupling. 
This yields a large number of excitation energies, preventing a simple peak assignment. 
The lowest peak at 0.10\,eV reflects the energy scale of spin-orbit coupling $\zeta$ 
but, as shown below, also is sensitive to hopping. 
In inelastic neutron scattering on polycrystalline samples, magnetic modes 
have been observed at 70 and 90\,meV \cite{Chen19}. The peak at 0.26\,eV can be 
traced back to hopping and $\zeta$ (see below). 
The peaks above 0.5\,eV predominantly can be attributed to the interplay of 
the trigonal crystal field, Hund's coupling, and hopping.

RIXS is the ideal tool to probe the quasimolecular character of excitations, 
as mentioned in the introduction.
With the dimer axis parallel to $c$ and an intra-dimer distance $d$, 
the interference pattern is expected to show a period 
$l_0$\,=\,$c/d$\,=\,5.9 as a function of $l$.
The data in Fig.\ \ref{fig:spectra}\textbf{a} roughly cover the range from 
$l$\,=\,\mbox{$-2.9$} to 2. 
In particular the peak at 0.1\,eV  exhibits a pronounced, non-monotonic variation of the 
intensity as a function of $l$.
However, we additionally have to consider polarization effects.
This is illustrated in Fig.\ \ref{fig:spectra}\textbf{b}, which shows spectra measured 
on the (001) surface. Again, strong intensity changes are observed
as a function of $\theta$. 
Note, e.g., the different peak intensities for the two curves 
measured with $(-0.7\,\,0\,\,4.3)$ (dark red) and $(0.7\,\,0\,\,4.3)$ (blue)   
with the same value of $l$. This particular intensity change cannot be 
caused by the dimer interference but must originate from polarization effects. 
In general, it is not trivial to quantitatively disentangle polarization 
and interference effects. Due to the many-body character of the states, 
the RIXS intensity is the squared sum of several terms, 
which leads to a full mixing of these effects. However, further insights can be 
obtained via a careful comparison with theory, as discussed below.
\\

\noindent
\textbf{DISCUSSION}
\\
For a Ru dimer with four $t_{2g}$ holes, we focus on the intra-$t_{2g}$ excitations below 2\,eV.\@ 
On each of the two sites $i$\,=\,1 and 2, we have to consider spin-orbit coupling $\zeta$, 
trigonal crystal-field splitting $\Delta_{\rm trig}$, 
and Coulomb repulsion in terms of Hubbard $U$ and Hund's coupling $J_{\rm H}$. 
The trigonal crystal field splits the $t_{2g}$ manifold into $a_{1g}$ and 
$e_g^\pi$ orbitals. 
Intersite hopping is diagonal for $a_{1g}$ and $e_g^\pi$ 
orbitals and is parameterized by $t_{a_{1g}}$ and $f$\,=\,$t_{e_g^\pi}/t_{a_{1g}}.$ 
The Hamiltonian reads \cite{Revelli19,Li20,Magnaterra25trimer}
\begin{equation}
	H = \sum_{i} \left( H_{{\rm SOC}, i} + H_{\Delta, i} + H_{{\rm C}, i} \right) + H_{t}.
\end{equation}
For $U$, $J_{\rm H}$, and $\zeta$, the relevant parameter range is well established from 
previous results on strongly correlated ruthenates. The on-site Coulomb repulsion $U$ 
is typically found to be 2 to 2.5\,eV, $J_{\rm H}$ is reported between 0.25 and 0.35\,eV,
and results for $\zeta$ range from 0.08 to 0.15\,eV   
\cite{Streltsov15,Suzuki19,Gretarsson19,Sarte20,Takahashi21,Vergara22,Krajewska24,Yuan25}. 
In contrast, the crystal-field splitting $\Delta_{\rm trig}$ and the hopping parameters 
$t_{a_{1g}}$ and $t_{e_g^\pi}$ may vary strongly between different compounds. 
\\

\begin{figure}[t]
	\centering
	\includegraphics[width=\columnwidth]{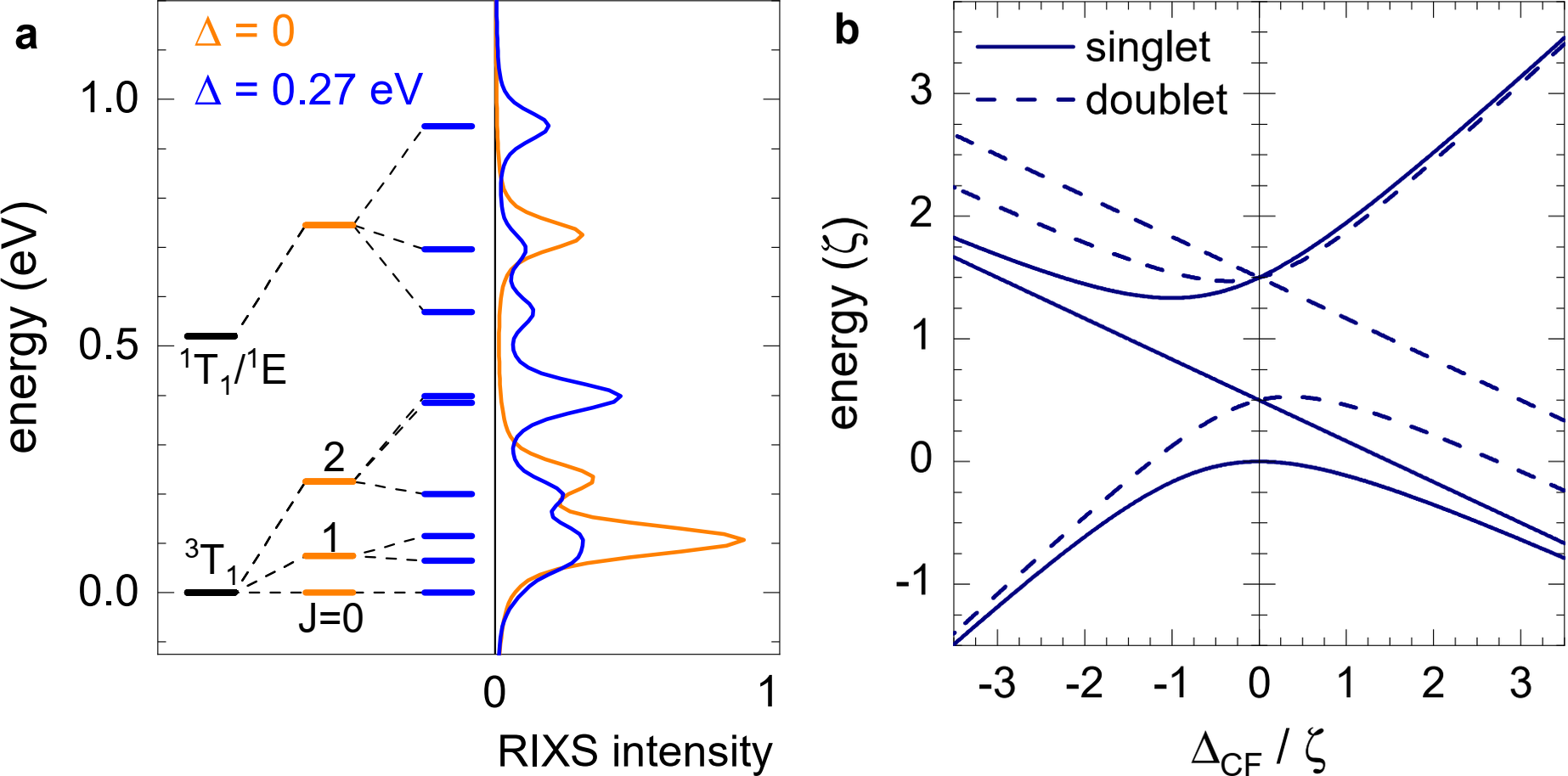}
	\caption{\textbf{Response of a single $t_{2g}^4$ site.} 
    \textbf{a} 
    The left panel shows the cubic multiplets (black) and how they are split by 
    spin-orbit coupling (orange) and a trigonal crystal field (blue). 
    The right panel depicts the corresponding calculated RIXS spectra for 
    $\zeta$\,=\,0.15\,eV, $J_{\rm H}$\,=\,0.26\,eV, and $\Delta_{\rm trig}$\,=\,0 or 0.27\,eV.\@ 
    \textbf{b} Energies within the $^3T_1$ manifold as a function of $\Delta_{\rm trig}/\zeta$ 
    in the limit of large $J_{\rm H}$.
    For $\Delta_{\rm trig}$\,=\,0, the states at 0, $\zeta/2$, and $3\zeta/2$ correspond 
    to $J$\,=\,0, 1, and 2, respectively.
}
\label{fig:single_site}
\end{figure}

\noindent
\textbf{Individual $4d^4$ Ru sites}
\\
We first address the electronic states of a single $4d^4$ Ru site, i.e., a site with 
two $t_{2g}$ holes, providing a suitable starting point for the discussion of a dimer. 
In cubic symmetry and for $\zeta$\,=\,0, Coulomb interactions lift the degeneracy of 
the $t_{2g}^4$ states, 
giving rise to a $^3T_1$ ground state and excitations at $2J_{\rm H}$ ($^1T_1$, $^1E$) and  
$5J_{\rm H}$ ($^1\!A_1$) \cite{Zhang17}. 
Spin-orbit coupling splits the $^3T_1$ multiplet into a $J$\,=\,0 
ground state and the $J$\,=\,1 and 2 excited states at $\zeta/2$ and $3\zeta/2$,  
see Fig.\ \ref{fig:single_site}\textbf{a}. 
In the ruthenates, one finds $J_{\rm H}/\zeta$\,$\approx$\,2 to 3, 
such that the two types of 
excitations with energies $\propto$\,$\zeta$ and $\propto$\,$J_{\rm H}$ are well separated 
in cubic compounds such as $4d^4$ K$_2$RuCl$_6$ \cite{Takahashi21}. 
The equivalent intra-$t_{2g}$ excitations also have been observed in, e.g., RIXS on cubic 
$5d^4$ K$_2$OsCl$_6$ \cite{Warzanowski23}. 
The RIXS intensity of excitations from $J$\,=\,0 
to the $^1\!A_1$ multiplet at $5J_{\rm H}$ vanishes 
for a scattering angle of 90$^\circ$ \cite{Warzanowski23}, as used in our experiment. 

In the ruthenates, the non-cubic crystal field splitting often is larger than $\zeta$. 
A large crystal field splits the multiplets at $2J_{\rm H}$ as well as the 
ninefold degenerate $^3T_1$ multiplet. Combined with spin-orbit coupling, this 
gives rise to a rich behavior at low energies, see Fig.\ \ref{fig:single_site}. 
With RIXS, the corresponding excitations have been studied in $4d^4$ Ca$_2$RuO$_4$, 
showing four peaks at about 0.05, 0.32, 0.75, and 1.0\,eV \cite{Gretarsson19}. 
Roughly, the lower two can be assigned to spin-orbit coupling and crystal-field splitting, 
while the two peaks at 0.75 and 1.0\,eV correspond to the feature at $2J_{\rm H}$ 
split by the crystal field, cf.\ Fig.\ \ref{fig:single_site}. 
A similar case has been reported in RIXS on $4d^4$ In$_2$Ru$_2$O$_7$ at 300\,K, 
showing five peaks at about 0.05, 0.28, 0.39, 0.70, and 1.0\,eV \cite{Krajewska24}. 
\\

\noindent
\textbf{Excitations on a dimer}
\\
In order to determine the electronic parameters, in particular $t_{a_{1g}}$, $t_{e_g^\pi}$, 
and $\Delta_{\rm trig}$, we numerically simulated the RIXS spectra, 
including the pronounced dependence on the scattering geometry, i.e., the sample orientation 
and angle of incidence $\theta$. The latter determines both $\mathbf{q}$ and the polarization. 
We used least error fitting to determine the relevant parameter regime, 
and then further optimized parameters in a narrow range.
The calculated spectra reproduce the key characteristics of the 
experimental data, see Fig.\ \ref{fig:spectra}.
The simulations describe the overall peak structure with two dominant peaks 
below 0.4\,eV and four main RIXS features above 0.5\,eV.\@ 
The main shortcoming is that the energies of the two lowest peaks are slightly too high 
in the calculations. 
However, the chosen parameter set considers the peak energies, the overall line shape, 
and the $\theta$ dependence of the intensity. We in particular achieve a good description 
of the latter, showing opposite behavior at low and high energies and for the two sample 
orientations. On the (110) surface, Fig.\ \ref{fig:spectra}\textbf{a}, 
the intensity is maximized for small $\theta$ (blue curve; positive $l$) 
below about 0.4\,eV but for large $\theta$ (dark red; negative $l$) at higher energies,
giving rise to a kind of isosbestic point at 0.4\,eV  where the RIXS intensity is 
nearly independent of $\theta$. The data on the (001) surface show the opposite behavior, 
with the intensity at low energy being maximized for large $\theta$ 
(dark red curve; negative $h$).
These features are very well reproduced by the simulation.

Optimal agreement between theory and experiment is obtained for 
$U$\,=\,2\,eV, $J_{\rm H}$\,=\,0.26\,eV, and $\zeta$\,=\,0.15\,eV.\@ 
These values are within the range established by previous studies on $4d$ ruthenates
\cite{Streltsov15,Suzuki19,Gretarsson19,Sarte20,Takahashi21,Vergara22,Krajewska24,Yuan25}. 
For the more material-specific parameters we find 
$\Delta_{\rm trig}$\,=\,0.27\,eV, $t_{a_{1g}}$\,=\,0.66\,eV, and $f$\,=\,$-0.45$.
A value of $f$ close to $-1/2$ agrees with theoretical predictions for face-sharing 
octahedra \cite{Li20}. Large hopping $t_{a_{1g}}$\,$\gtrsim$\,0.7\,eV has also been 
reported for Ba$_3$LaRu$_2$O$_9$ with five $t_{2g}$ holes per dimer \cite{Yuan25}. 
Density-functional theory predicts $t_{a_{1g}}$\,$\approx$\,0.4-0.8\,eV 
for face-sharing ruthenates \cite{Li20}.
In Ba$_3$CeRu$_2$O$_9$, the octahedra are elongated along the dimer axis, indicating 
a negative point-charge contribution to $\Delta_{\rm trig}$ in the hole picture, 
but a dominant covalent contribution may reverse the sign \cite{Kugel15, Khomskii16}.

\begin{figure}[t]
	\centering
	\includegraphics[width=\columnwidth]{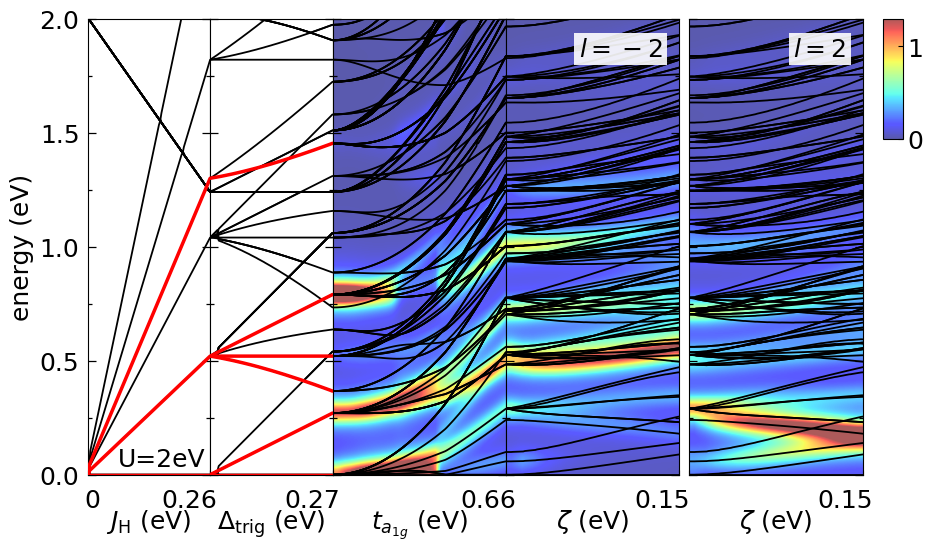}
	\caption{\textbf{Excitation energies and RIXS intensity of a face-sharing dimer 
    with four $t_{2g}$ holes.} 
    From left to right, different parameters are successively included. 
    We use $U$\,=\,2\,eV and first increase $J_{\rm H}$ from 0 to 0.26\,eV, 
    then $\Delta_{\rm trig}$ from 0 to 0.27\,eV, hopping $t_{a_{1g}}$ up to 0.66\,eV 
    with fixed $f$\,=\,$-0.45$, and finally $\zeta$ from 0 to 0.15\,eV.\@ 
    The red lines in the first two panels represent the energies of a single site with 
    two holes. 
    The color plot in the third and fourth panels shows the calculated RIXS intensity 
    for the (110) orientation with $2\theta$\,=\,90$^\circ$ and $l$\,=\,-2 
    for a peak width of 50\,meV.\@  
    The last panel again depicts the energies as a function of $\zeta$ but employs 
    $l$\,=\,2 for the RIXS intensity to highlight the behavior at low energies, 
    cf.\ Fig.\ \ref{fig:spectra}\textbf{c}.
}
\label{fig:energy_levels}
\end{figure}

For the peak assignment, Fig.\ \ref{fig:energy_levels} shows the 
excitation energies for the best parameter set.
In the left panel, we start with $U$\,=\,2\,eV and switch on $J_{\rm H}$ up to 0.26\,eV.\@ 
The two red lines denote the excitations of a single site at $2J_{\rm H}$ and $5J_{\rm H}$. 
Three further lines correspond to excitations on both sites with total excitation energies 
of $4J_{\rm H}$, $7J_{\rm H}$, and  $10J_{\rm H}$. For vanishing hopping, such double excitations  
have zero intensity in RIXS.\@ The latter is also valid for the excitation at $U-3J_{\rm H}$, 
i.e., with an energy that \textit{decreases} with increasing $J_{\rm H}$. 
It corresponds to the lowest intersite excitation, 
$d_1^4d_2^4$\,$\rightarrow$\,$d_1^3d_2^5$ \cite{Zhang17}. 
The second panel depicts the effect of varying $\Delta_{\rm trig}$ from 0 to 0.27\,eV.\@ 
The red lines again refer to a single site, showing the splitting of the cubic multiplets, 
see also Fig.\ \ref{fig:single_site}.
Finally, the third and fourth panel show the effects of hopping and spin-orbit coupling, 
respectively. The underlying color plot depicts the calculated RIXS intensity 
for $l$\,=\,-2. In contrast, the fifth panel employs $l$\,=\,2, highlighting the 
low-energy peaks, cf.\ Fig.\ \ref{fig:spectra}\textbf{c}.

Intra-dimer hopping substantially increases the number of distinct excitation energies.
Remarkably, most excitation energies increase with increasing hopping, showing 
that the ground-state energy $E_0$ is one of those that benefit the most. 
Concerning the ground state, we find a level crossing at $t_{a_{1g}}$\,$\approx$\,0.4\,eV, 
which is evident from a jump in the RIXS intensity, see Fig.\ \ref{fig:energy_levels}.
The character of the ground state will be discussed below.
Here we only mention that, for small $t_{a_{1g}}$, the energy is lowered $\propto t_{a_{1g}}^2/U$, 
as expected for a Mott insulator with exchange interactions. 
This picture breaks down due to level crossing. However, even for $t_{a_{1g}}$ around 0.66\,eV 
we find that $E_0$ is lowered roughly quadratically in hopping. The behavior strongly differs 
from the energy of a bonding state that decreases linearly in hopping in the fully delocalized 
quasimolecular limit. This reflects the rather localized character of the states, 
despite the large hopping and the breakdown of the exchange limit.

Based on the more localized character, the results for a single site discussed above 
to some extent provide a guideline for the 
interpretation of some of the RIXS peaks of Ba$_3$CeRu$_2$O$_9$. This works 
in particular at high energy and as long as the hopping-induced energy shifts 
are small compared to $2J_{\rm H}$, which is true for many but not all of the states.
The RIXS peaks at 0.8 and 1.1\,eV are related to the multiplets at $2J_{\rm H}$, 
split by $\Delta_{\rm trig}$ and shifted in energy by hopping, see Fig.\ \ref{fig:energy_levels}. 
The peak at 0.53\,eV predominantly can be traced back to the effect of $\Delta_{\rm trig}$ 
and hopping.
Remarkably, the three energies of 0.53, 0.80, and 1.1\,eV are roughly 0.1\,eV higher 
than the peak energies reported for single-site $4d^4$ Ca$_2$RuO$_4$ and 
In$_2$Ru$_2$O$_7$ \cite{Gretarsson19,Krajewska24}.
This may indicate a common origin, where the energy shift in Ba$_3$CeRu$_2$O$_9$ 
is caused by the hopping-induced lowering of the ground state. 

The low-energy peak at 0.1\,eV can be attributed to spin-orbit coupling, 
see Fig.\ \ref{fig:energy_levels}, which again to some extent is reminiscent 
of Ca$_2$RuO$_4$ and In$_2$Ru$_2$O$_7$ \cite{Gretarsson19,Krajewska24}.
In Ba$_3$CeRu$_2$O$_9$, however, one has to address the effect of spin-orbit coupling on the 
low-energy dimer states, for which hopping is essential, as discussed below. 
Finally, excitations around 0.26\,eV mainly arise due to hopping, 
but also spin-orbit coupling plays a role, see Fig. \ref{fig:energy_levels}.
\\

\noindent
\textbf{Intensity modulation}
\\
The excellent description of the $\theta$ dependence of the RIXS intensity is a strong 
asset of our theoretical result. The angle of incidence $\theta$ sets polarization and 
$\mathbf{q}$, and both affect the intensity. 
For RIXS on a dimer with quasimolecular states, a given excited state 
can be reached by scattering on either of the two sites. Summation over the coherent 
scattering processes yields a sinusoidal intensity modulation \cite{Revelli19}. 
With inversion symmetry, one expects either $\sin^2(\pi l/l_0)$ or $\cos^2(\pi l/l_0)$ 
behavior (with $l_0$\,=\,$c/d$), depending on the parity of the involved states. 
The excitation from an, e.g., even ground state to an even excited state with identical 
matrix elements on both sites yields a $\cos^2(\pi l/l_0)$ modulation. 
A face-sharing dimer does not obey inversion symmetry, but the crystal structure shows two 
dimer orientations that are rotated by $\pi$ around $c$, see Fig.\ \ref{fig:structure}. 
Summing over both orientations again yields a $\sin^2(\pi l/l_0)$ or $\cos^2(\pi l/l_0)$ 
interference pattern \cite{Revelli19}.

In a Mott insulator, the picture is different. For a strictly local excitation 
on site $i$, the RIXS intensity does not depend on $\mathbf{q}$ \cite{Magnaterra23Ti}. 
Orbital excitations typically are considered to be such local excitations, e.g., 
from $|xy\rangle_i$ to $|yz\rangle_i$. 
In this example, exchange coupling between the two dimer sites will yield states 
$|yz\rangle_1 \pm |yz\rangle_2$ 
but this will only cause a modulation if the energy 
separation is larger than the peak width. Note that a large energy splitting marks the 
crossover to the quasimolecular cluster Mott case. 
In the Mott limit, the superposition of overlapping $\sin^2(\pi l/l_0)$ and 
$\cos^2(\pi l/l_0)$ modulations gives constant intensity as a function of $\mathbf{q}$. 
The situation is different for excitations between entangled states, which in the Mott limit 
typically is the case for spin excitations. For simplicity, we consider two sites 
carrying $S$\,=\,1/2 each. The excitation from a singlet 
$(|\!\!\uparrow \downarrow\rangle - |\!\!\downarrow\uparrow\rangle)/\sqrt{2}$ to 
a triplet state $|\!\!\uparrow \uparrow\rangle$
can be reached by a spin flip on either of the two sites. This again yields a sinusoidal 
intensity modulation, as observed for the bond-directional magnetic excitations in the 
Kitaev material Na$_2$IrO$_3$ \cite{Revelli20,Magnaterra23}.

\begin{figure}[t]
	\centering
	\includegraphics[width=\columnwidth]{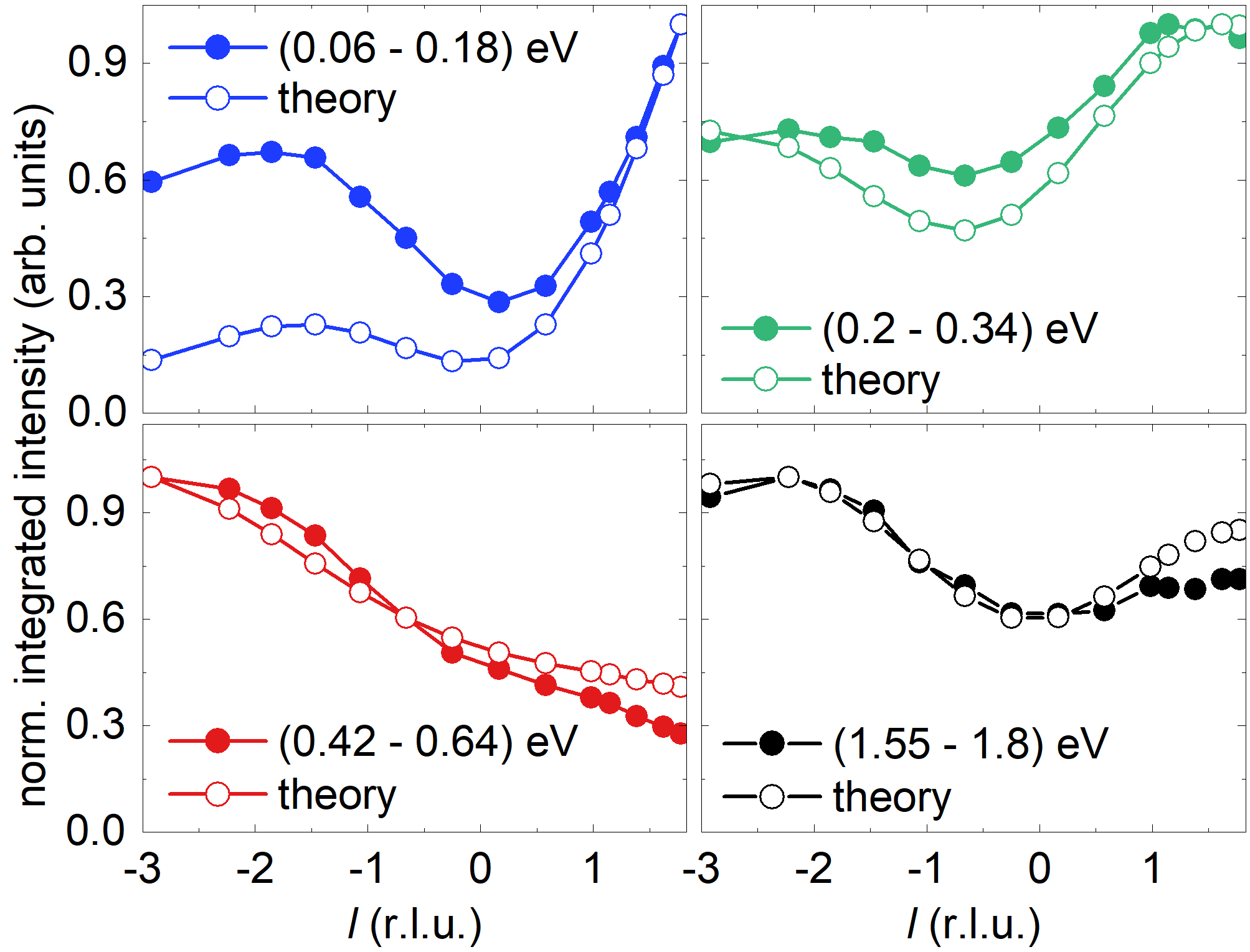}
	\caption{\textbf{RIXS intensity as a function of $l$.}
    The experimental data (full symbols) have been integrated over the indicated 
    energy intervals and normalized to the maximum value. 
    The corresponding simulations have been evaluated over the respective peak regions. 
    The calculations reproduce the overall behavior of the experimental result very well.
}
\label{fig:integrals}
\end{figure}

In both the Mott limit or the cluster Mott limit, a possible modulation will show minimum 
or maximum intensity for $l$\,=\,0, which is covered by the (110) orientation. 
The experimental data indeed show minimum intensity close to $l$\,=\,0 below 0.35\,eV 
as well as around 1.7\,eV.\@ 
This is highlighted in Fig.\ \ref{fig:integrals}, which shows the RIXS intensity integrated 
over selected energy ranges, both for experiment and theory.
Below 0.35\,eV (blue and green) and above 1.55\,eV (black), 
the integrated intensity clearly shows non-monotonic behavior, 
and the modulation agrees with a $\sin^2(\pi l/l_0)$ behavior that acquires asymmetry 
with respect to $l$\,=\,0 due to polarization effects. 

In contrast, we find a monotonic decrease of intensity around, e.g., 0.5\,eV, 
which we attribute to dominant polarization effects. These arise because a change 
of $\mathbf{q}$ is accompanied by a change of the actual scattering geometry, i.e., $\theta$.
This assignment is supported by simulations for a single $4d^4$ site with identical 
parameters, in particular positive $\Delta_{\rm trig}$ but vanishing hopping. 
We find a very similar monotonic trend for the $\theta$ dependence with opposite behavior 
at low and high energies and for the two sample orientations, see App.\ \ref{sec:App_RIXS_single}.

In the $5d$ iridate dimers Ba$_3$$M$Ir$_2$O$_9$, the entire intra-$t_{2g}$ excitation spectrum 
shows strong modulation of the RIXS intensity \cite{Revelli19,Revelli22,Magnaterra23Ti}. 
The ruthenate Ba$_3$CeRu$_2$O$_9$ shows a different behavior. 
The modulation for many peaks is suppressed or overruled by 
polarization effects, pointing to a more local character. However, this may also be 
caused by averaging over the large number of excitations of the four-hole dimer, 
see Fig.\ \ref{fig:energy_levels}. 
Finally, the observation of a pronounced modulation as a function of $l$ 
both at low and high energies supports a partially quasimolecular character. 
Overall, the intensity modulation supports the picture of Ba$_3$CeRu$_2$O$_9$ being 
located in the intermediate regime. 
However, to quantitatively understand the puzzling character of the ground state, 
we have to address the wavefunction obtained in our simulations of the RIXS data.
\\

\noindent
\textbf{Character of electronic states}
\\
For the four-hole Ru dimers, the competition of Coulomb repulsion, hopping, crystal-field splitting, 
and spin-orbit coupling allows for several distinct ground states. In the following, we discuss 
their character and compare in particular the local Mott limit and the quasimolecular limit. 
We substantiate the claim that Ba$_3$CeRu$_2$O$_9$ is best described as being in the 
intermediate regime.

\begin{figure}[t]
	\centering
	\includegraphics[width=\columnwidth]{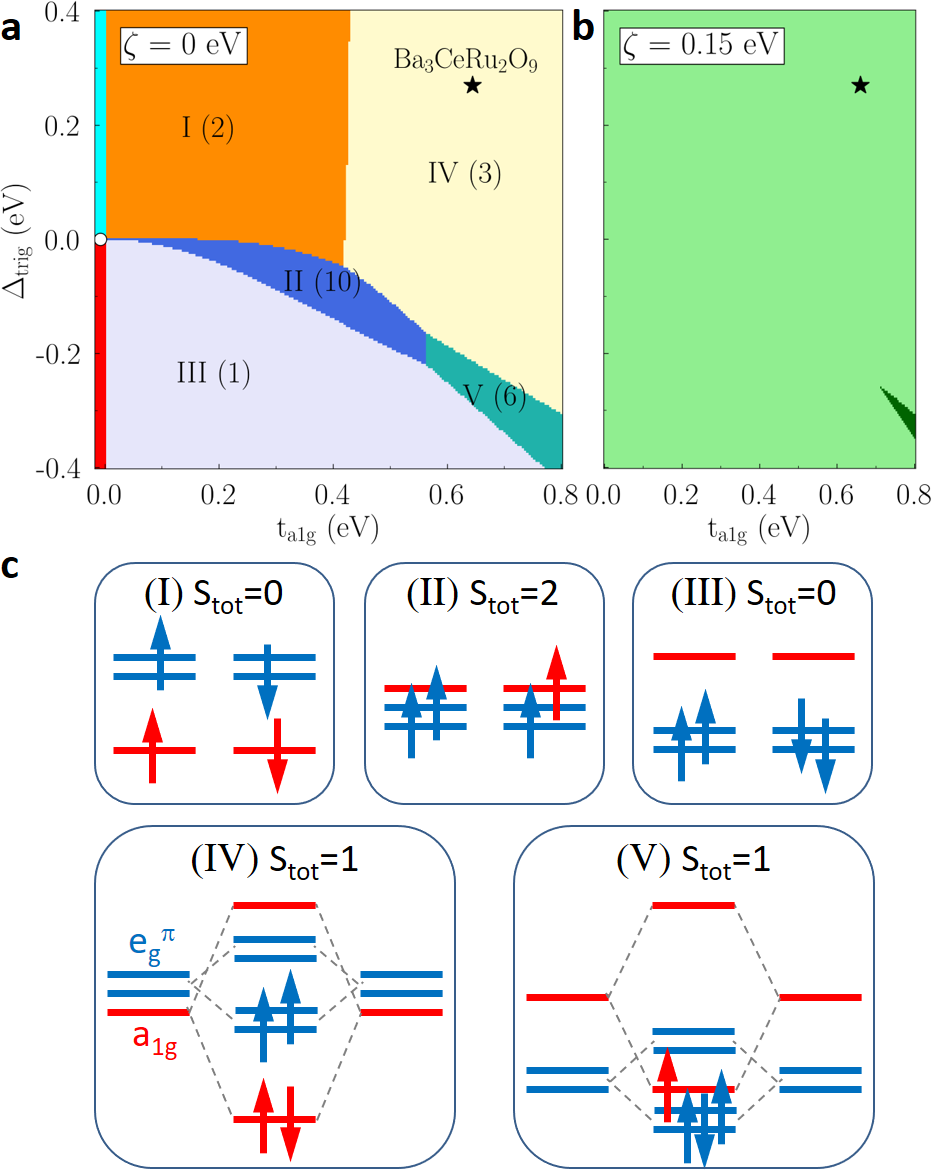}
	\caption{\textbf{Ground states of a four-hole dimer in different limits.}
	\textbf{a} For $\zeta$\,=\,0, 
    we find several ground states as a function of 
    $\Delta_{\rm trig}$ and $t_{a_{1g}}$. The dominant states I-V for $t_{a_{1g}}$\,$>$\,0 
    are plotted for $U$\,=\,2\,eV, 
    $J_{\rm H}$\,=\,0.26\,eV, and $t_{e_g^\pi}/t_{a_{1g}}$\,=\,$-0.45$. 
    The numbers in parentheses give the degeneracy.
    Note that there are several tiny pockets of further phases at some parts of the phase boundaries 
    that are not resolved in the figure and are irrelevant for our discussion. 
    Additionally, the three states for $t_{a_{1g}}$\,=\,0 and positive, vanishing, or 
    negative $\Delta_{\rm trig}$ are indicated on the very left. 
	  \textbf{b} Phase diagram for $\zeta$\,=\,0.15\,eV.\@ 
    A spin-orbital singlet state (light green) dominates, and the parameter region with 
    orbital degeneracy (dark green) has shrunken considerably.
    \textbf{c} The sketches for I-III depict local $a_{1g}$ (red) and $e_g^\pi$ (blue) orbitals, 
    while IV and V also show \mbox{(anti-)} bonding orbitals for the limit of large hopping. 
} 
\label{fig:sketch}
\end{figure}

\textbf{Mott limit for $\zeta$\,=\,0:} 
We start from $\zeta$\,=\,0, a case that is often addressed for ruthenates and 
other $4d$ compounds \cite{Streltsov16}, also in previous reports on 
Ba$_3$CeRu$_2$O$_9$ \cite{Doi01,Chen19}.
Indeed we find that the results for $\zeta$\,=\,0 are helpful for understanding 
the physics for finite $\zeta$.
In the localized Mott limit, Coulomb repulsion suppresses charge fluctuations such that 
each Ru site hosts two $t_{2g}$ holes. Hund's coupling then favors local $S$\,=\,1 configurations. 
The resulting dimer ground state depends sensitively on the trigonal crystal-field splitting 
$\Delta_{\mathrm{trig}}$, which controls the orbital occupancy, and on the 
strength of hopping.
The corresponding phase diagram is summarized in Fig.\ \ref{fig:sketch}\textbf{a} 
for $U$\,=\,2\,eV, $J_{\rm H}$\,=\,0.26\,eV, and $f$\,=\,$-0.45$.

For $t_{a_{1g}}$\,=\,0, each Ru site shows 3-fold spin degeneracy. 
The orbital degeneracy depends on $\Delta_{\rm{trig}}$, which yields dimer ground states 
with total degeneracy of either 
36, 81, or 9 for $\Delta_{\rm trig}$ being positive, zero, or negative, respectively. 
Small hopping $t_{a_{1g}}$ causes exchange interactions between the $S$\,=\,1 sites. 
Both antiferromagnetic and ferromagnetic coupling can be realized, yielding 
$S_{\rm tot}$\,=\,0 in states I and III but $S_{\rm tot}$\,=\,2 in state II.\@ 
In I for $\Delta_{\rm{trig}}$\,$>$\,0, one hole per site occupies the $a_{1g}$ orbital 
and the second one an $e_g^\pi$ orbital, see Fig.\ \ref{fig:sketch}\textbf{c}. 
In this situation, hopping $t_{e_g^\pi}$ between the degenerate $e_g^\pi$ orbitals yields 
a Kugel-Khomskii-type exchange that favors parallel spin alignment and the occupation of 
different orbitals. 
However, this is overruled by the stronger $t_{a_{1g}}$ that favors antiparallel spin alignment. 
Altogether, only a two-fold orbital degeneracy remains in state I with $S_{\rm tot}$\,=\,0.
In state III for $\Delta_{\rm{trig}}$\,$<$\,0, both holes preferentially occupy 
the $e_g^\pi$ sector, so that the orbital degeneracy is removed already at the local level, 
see Fig.\ \ref{fig:sketch}\textbf{c}.
In the weak-hopping limit, antiferromagnetic exchange yields $S_{\rm tot}$\,=\,0 for III.\@

In contrast, state II is realized for small $\Delta_{\rm trig}$ with nearly degenerate orbitals.
In this case, the larger $a_{1g}$ hopping selects configurations with in total one $a_{1g}$ hole, 
while Hund's coupling favors parallel spins in the virtual intermediate states. 
As a result, the effective interaction between the two sites is ferromagnetic and the dimer realizes 
a high-spin state with $S_{\rm tot}$\,=\,2. 
This may be viewed as a form of double exchange \cite{Streltsov16,Khomskii21}, 
in which the strongly hopping $a_{1g}$ hole mediates ferromagnetic coupling between 
the more localized $e_g^\pi$ degrees of freedom. 
Because the $e_g^\pi$ orbitals remain degenerate, state II carries a 2-fold orbital degeneracy 
on top of the 5-fold spin degeneracy.

\textbf{Quasimolecular limit for $\zeta$\,=\,0:}
In the opposite limit of dominant intra-dimer hopping, the most appropriate description is 
in terms of bonding and antibonding quasimolecular orbitals. 
Because $|t_{a_{1g}}| > |t_{e_g^\pi}|$, the bonding $a_{1g}$ orbital is filled first, 
while the remaining two holes occupy the bonding $e_g^\pi$ sector. 
Hund's coupling then favors a triplet with $S_{\rm tot}$\,=\,1, see state IV 
in Fig.\ \ref{fig:sketch}\textbf{c}.
For $\Delta_{\mathrm{trig}}$\,$>$\,0, this regime is reached already for moderate hopping 
$t_{a_{1g}}$\,$\approx$\,0.4\,eV, 
where the ground state exhibits substantial localized character, 
see Fig.\ \ref{fig:sketch}\textbf{a}.
However, it is continuously connected to the quasimolecular limit.
This phase extends to increasingly negative values of $\Delta_{\rm trig}$ 
as the hopping increases. 
This can be understood naturally from the quasimolecular perspective: 
because $|t_{a_{1g}}| > |t_{e_g^\pi}|$, the energy gain associated with occupying the 
bonding $a_{1g}$ orbital eventually outweighs the crystal-field energy cost incurred for 
$\Delta_{\rm trig}$\,$<$\,0. 
However, for a given value of hopping, large negative $\Delta_{\rm trig}$ causes a transfer 
of holes from bonding $a_{1g}$ to bonding $e_g^\pi$, leaving either one or zero $a_{1g}$ holes 
in states V and III, respectively.

Above we have identified the main character of the RIXS peaks and revealed 
and described the \textbf{q}-dependent modulation of the RIXS intensity.
Our ED simulations reproduce the main experimental RIXS features very well 
and constrain the relevant parameter regime. 
Neglecting small $\zeta$, Ba$_3$CeRu$_2$O$_9$ lies well within the range of state IV, 
see star in Fig.\ \ref{fig:sketch}\textbf{a}. On the one hand, about 74\,\% of the 
ground state wavefunction belongs to the localized limit with two holes per site. 
On the other hand, state IV is well understood in the quasimolecular limit 
but cannot be rationalized in the weak-coupling limit. 
Compared to state I, it requires a hopping of sufficient size to violate Hund's rule 
and obtain $S_{\rm tot}$\,=\,1. This clearly demonstrates the intermediate character 
of Ba$_3$CeRu$_2$O$_9$.
\\

\noindent
\textbf{Nature of the ground state for finite $\zeta$:}
\\
The phase diagram for finite $\zeta$ looks much simpler, see Fig.\ \ref{fig:sketch}\textbf{b}.
However, finite $\zeta$ does not qualitatively invalidate our classification for $\zeta$\,=\,0. 
Spin-orbit coupling removes some of the exact degeneracies and turns the level crossings of the 
$\zeta$\,=\,0 phase diagram into avoided crossings. The main character of the ground state 
nevertheless changes as a function of hopping and $\Delta_{\rm trig}$, and the analysis 
for $\zeta$\,=\,0 provides a useful guide. In particular, it remains valid that Ba$_3$CeRu$_2$O$_9$ 
lies outside the weak-coupling regime of simple exchange between localized states. 
To gain further insight, we approximate the ED ground state by simple trial wave functions.

\textbf{Mott limit for finite $\zeta$:}
About 74\,\% of the ground-state weight resides in configurations with two holes 
on each Ru site. This already indicates a predominantly localized character
and motivates a description in terms of local building blocks.
Finite $\zeta$ lifts the degeneracy within the $e_g^\pi$ sector. 
Using the complex orbitals $e_{g\pm}^\pi$ is the most convenient choice, 
as $\zeta$ merely shifts $|e_{g+}^\pi,\uparrow\rangle$ and $|e_{g-}^\pi,\downarrow\rangle$ 
upwards in energy without mixing with the $a_{1g}$ orbitals. 
In good approximation, we may restrict the discussion to 
\begin{align}
\nonumber
    |a_{1g},\sigma\rangle \equiv |a\sigma\rangle , & &
    |e_{g+}^\pi,\downarrow\rangle \equiv |+\!\downarrow\rangle  , & &
    |e_{g-}^\pi,\uparrow\rangle \equiv |-\!\uparrow\rangle  \, . 
\end{align}
The definitions of the orbitals and their relation to the spin-orbit eigenstates 
$|j,j_z\rangle$ are given in Appendix \ref{sec:jzEigenstates}.
Using this local basis, we define the singlet states
\begin{align}
    |\psi_1\rangle &= \Big[|a\!\uparrow\rangle_1 |+\!\downarrow\rangle_1 
       |a\!\downarrow\rangle_2 |-\!\uparrow\rangle_2 + (1 \leftrightarrow 2 )\Big]/\sqrt{2} 
       \nonumber \\
	|\psi_2\rangle &= \Big[|a\!\uparrow\rangle_1 |-\!\uparrow\rangle_1 
       |a\!\downarrow\rangle_2 |+\!\downarrow\rangle_2 + (1 \leftrightarrow 2 )\Big]/\sqrt{2} \nonumber\\
    |\psi_3\rangle & = \Big[
	|a\!\uparrow\rangle_1 |a\!\downarrow\rangle_1 |-\!\uparrow\rangle_1 |+\!\downarrow\rangle_2  
%    \nonumber \\ & 
    +|a\!\uparrow\rangle_1 |a\!\downarrow\rangle_1 |-\!\uparrow\rangle_2 |+\!\downarrow\rangle_1 
 \nonumber \\ 	& 
      -|a\!\uparrow\rangle_1 |a\!\downarrow\rangle_2 |-\!\uparrow\rangle_1 |+\!\downarrow \rangle_1 
%    \nonumber\\ & 
      -|a\!\uparrow\rangle_2 |a\!\downarrow\rangle_1 |-\!\uparrow\rangle_1 |+\!\downarrow\rangle_1 
      \nonumber\\ &  + (1 \leftrightarrow 2) \Big]/\sqrt{8} \, . 
\end{align}
With the constraint $|\alpha|^2+|\beta|^2+|\gamma|^2=1$, the trial state
\begin{equation}
\label{eq:psiMott}
	|\psi\rangle = \alpha |\psi_1\rangle + \beta |\psi_2\rangle + \gamma |\psi_3\rangle
\end{equation}
captures more than 86\,\% of the ED ground state using only two independent parameters. 
Here, $|\psi_1\rangle$ provides the dominant contribution within the sector with two holes 
per site ($\approx$\,46\,\%). Note that $|\psi_1\rangle$ violates Hund's rule, showing 
that it lies outside the weak-coupling exchange limit given by states I-III.\@ 
In contrast, the smaller contribution $|\psi_2\rangle$ is connected to state I.\@ 
The leading correction with asymmetric charge distribution is given by $|\psi_3\rangle$ 
carrying 24\,\% of the ED ground state.

\textbf{Quasimolecular limit for finite $\zeta$:}
In agreement with the intermediate nature of Ba$_3$CeRu$_2$O$_9$, the ground state 
can similarly be approximated in a cluster Mott picture.
Somewhat surprisingly, a trial state based on the quasimolecular limit performs even slightly better than the Mott limit one in Eq.\ \eqref{eq:psiMott}.
The simple product state
\begin{equation}\label{eq:tilde_psi_0}
	|\tilde{\psi}_0\rangle
	= |a\!\uparrow\rangle_B \,
	|a\!\downarrow\rangle_B \,
	|-\!\uparrow\rangle_B \,
	|+\!\downarrow\rangle_B,
\end{equation}
where $|\alpha\rangle_B$ denotes the bonding state of state $|\alpha\rangle$, 
already captures  71\,\%  of the ED ground state. 

The quasimolecular ansatz can be systematically improved by incorporating the effect of Coulomb repulsion, which enhances the weight of configurations with two holes per site and suppresses sectors with three or four holes on one site.
Within the (anti-)bonding basis, this can, e.g., be achieved by admixing states with 
an even number of antibonding (AB) orbitals. 
Using
\begin{align}\label{eq:bonding}
	|\tilde\psi_2\rangle & = \frac{1}{\sqrt{6}}\Big[ |a\!\uparrow\rangle_{AB}\,\, 
	|a\!\downarrow\rangle_{AB}\,\,
	|-\!\uparrow\rangle_B\,\,
	|+\!\downarrow\rangle_B \,  
    \nonumber\\
	& + |a\!\uparrow\rangle_{AB}\,\, 
    |-\!\uparrow\rangle_{AB}\,\,
	|a\!\downarrow\rangle_{B}\,\,
	|+\!\downarrow\rangle_B \, +\ldots \Big] \, ,
    \nonumber\\
	|\tilde\psi_4\rangle &=  |a\!\uparrow\rangle_{AB}\,\, 
	|a\!\downarrow\rangle_{AB}\,\,
	|-\!\uparrow\rangle_{AB}\,\,
	|+\!\downarrow\rangle_{AB} \, ,
\end{align}
 the trial state
\begin{equation}
	|\tilde{\psi}\rangle
	= \alpha |\tilde{\psi}_0\rangle
	+ \beta |\tilde{\psi}_2\rangle
	+ \gamma |\tilde{\psi}_4\rangle .
\end{equation}
captures  87\,\% of the ED ground state, again with only two independent parameters.
Remarkably, we find a nearly as good description of the ED ground state by replacing the bonding 
and antibonding combinations of
$\{|a\!\uparrow\rangle$, $|a\!\downarrow\rangle$, $|-\!\uparrow\rangle$, $|+\!\downarrow\rangle\}$
by those constructed from the $j$ eigenstates
$\{|\tfrac{1}{2},\tfrac{1}{2}\rangle, |\tfrac{1}{2},-\tfrac{1}{2}\rangle, |\tfrac{3}{2},\tfrac{1}{2}\rangle, |\tfrac{3}{2},-\tfrac{1}{2}\rangle\}$, see App.\ \ref{sec:App_states} for details.
In face-sharing iridate dimers and trimers with large spin-orbit coupling $\zeta$\,$\approx$\,0.4\,eV, 
the quasimolecular $j$ states provide the most appropriate basis \cite{Revelli19,Revelli22,Magnaterra25trimer}. 
It is astounding how well the quasimolecular $j$ basis works even in Ba$_3$CeRu$_2$O$_9$, 
despite the much smaller value of $\zeta$ and the mainly localized nature of the ground state.

Overall, both the localized and quasimolecular constructions capture substantial fractions 
of the ground state. This again shows that Ba$_3$CeRu$_2$O$_9$ lies in the crossover regime 
between localized and quasimolecular behavior.
\\

\noindent
\textbf{CONCLUSIONS}
\\
The seemingly harmless non-magnetic dimer ground state of Ba$_3$CeRu$_2$O$_9$ with four holes 
per dimer turns out to be not trivial. It is located in the intriguing crossover regime 
between the local Mott limit and the quasimolecular cluster Mott limit. 
The charge distribution predominantly follows the expectations for strong Coulomb repulsion 
favoring two holes per site. However, hopping is so large that the ground state cannot be described 
in the weak-coupling limit with exchange interactions but it can be motivated from a quasimolecular 
perspective. Moreover, a simple quasimolecular trial wave function describes the ground state 
very well. 
Our results reveal the limits of the often considered dichotomy 
between localized states and quasimolecular states and highlight the more 
subtle physics of the crossover regime.

In cluster Mott insulators, the character of the quasimolecular 
magnetic moments is sensitive to electronic parameters \cite{Magnaterra24}. 
For instance the $5d$ iridate Ba$_3$InIr$_2$O$_9$ with three holes per 
dimer is close to the transition between $j$\,=\,3/2 and 1/2, governed 
by the size of hopping \cite{Li20,Revelli22}. In comparison, the 
crossover regime realized in the ruthenate Ba$_3$CeRu$_2$O$_9$ offers 
a larger variety of competing ground states as a function of hopping and 
trigonal crystal field, which are the electronic parameters that can be tuned most 
directly via, e.g., chemical pressure or external pressure. 
It is promising to extend our analysis to Ru dimer compounds with an odd number 
of holes that carry a local magnetic moment \cite{Li20}. In some cases, the classification 
as Mott insulators vs.\ quasimolecular compounds may have to be revisited. 
Indeed, a strong sensitivity to small changes of the crystal structure has been reported for 
Ba$_3$$M$Ru$_2$O$_9$ with $M^{3+}$ ions \cite{Yuan25}.

Spin-orbit coupling is the smallest of the electronic parameters 
and has often been neglected in $4d$ ruthenates. 
We find that comparing $\zeta$\,=\,0 and finite $\zeta$ is most helpful 
for understanding the rich many-body behavior.
Finite $\zeta$\,=\,0.15\,eV is decisive for the non-magnetic singlet 
ground state which the four-hole dimer features in an overwhelming part 
of the phase diagram. This bears some analogy with the 
local $J$\,=\,0 ground state of a single $d^4$ site that is realized irrespective of the value of 
$\Delta_{\rm trig}/\zeta$. For a lattice of $d^4$ sites with large $\Delta_{\rm trig}$, however, 
the local $J$\,=\,0 state does not necessarily provide the most intuitive picture 
compared to, e.g., an effective low-energy $S$\,=\,1 model. 
Similarly, the picture of a nonmagnetic dimer misses the underlying sensitivity of the 
electronic structure to the tight competition of several electronic parameters, 
giving rise to a multitude of possible ground states in a $\zeta$\,=\,0 approach
and to many low-energy states.
\\

\noindent
\textbf{METHODS}
\\

\noindent
\textbf{Crystal growth and crystal structure}
\\
We studied single crystals of Ba$_3$CeRu$_2$O$_9$.
Initially, polycrystalline Ba$_3$CeRu$_2$O$_9$ has been synthesized via conventional 
solid-state reactions, similar to previous reports \cite{Doi01}. The starting materials 
BaCO$_3$, CeO$_2$, and RuO$_2$ were weighed in appropriate metal ratios, thoroughly mixed 
and heated in alumina crucibles at 1573\,K for 48 hours. Phase purity was examined 
by powder x-ray diffraction, which beyond Ba$_3$CeRu$_2$O$_9$ revealed minor secondary phases 
of Ba$_4$CeRu$_3$O$_{12}$ and BaCeO$_3$.
Single crystals were subsequently grown using a flux method inspired by 
Refs.\ \cite{Revelli22, Revelli19, Magnaterra23Ti, Thakur20}. The prereacted polycrystalline 
powder was mixed with BaCl$_2$ $\cdot$ 2H$_2$O in a 1:30 molar ratio and heated in alumina 
crucibles to 1573\,K.\@ The melt was slowly cooled to 1173\,K at a rate of 2\,K/h. 
After cooling, residual BaCl$_2$ flux was dissolved using distilled water. 
For the resulting crystals, good agreement with the nominal stoichiometry was verified 
by energy-dispersive x-ray spectroscopy.

At 300\,K, Ba$_3$CeRu$_2$O$_9$ exhibits hexagonal symmetry (space group \textit{P6$_3$/mmc}) 
with lattice constants $a$\,=\,5.8878\,\AA{} and $c$\,=\,14.644\,\AA{} 
and intra-dimer Ru-Ru distance $d$\,=\,2.48\,\AA, see Fig.\ \ref{fig:structure}.
We studied single crystals with hexagonal shape and an area of about 0.45\,mm $\times$ 0.4\,mm 
perpendicular to the $c$ axis and a thickness of roughly 0.15\,mm along $c$. 
\\

\noindent
\textbf{RIXS}
\\
We performed RIXS measurements at the Ru \textit{L$_3$} edge in horizontal scattering 
geometry at beamline P01 at PETRA-III \cite{Gretarsson20}. 
The incoming x-rays were first monochromatized by a pair of cryogenically cooled asymmetric 
Si(111) crystals to obtain a bandwidth of about 0.6\,eV.\@ A secondary four-bounce monochromator 
(asymmetric) further reduced the bandwidth to about 60\,meV. We achieved an excellent total 
energy resolution of 64\,meV by using a SiO$_2$ $(10\bar{2})$ diced analyzer crystal with 
a rectangular mask of 30\,mm height. 
We measured at 20\,K with incident $\pi$ polarization and a fixed scattering angle of 
$2\theta$\,=\,90$^\circ$, strongly suppressing elastic (Thomson) scattering. 
For each angle of incidence $\theta$, the energy of zero loss was determined by measuring 
elastic scattering from GE varnish applied next to the sample. All RIXS data have been corrected 
for self-absorption effects based on the scattering geometry and the energy of the scattered 
photons \cite{Minola15}, using an x-ray absorption spectrum measured at 
2$\theta$\,=\,90$^{\circ}$ and $\theta$\,=\,45$^{\circ}$ on the (001) facet.

Concerning RIXS interferometry, the fixed scattering angle of 90$^\circ$ yields a fixed modulus 
$|\mathbf{q}|$ such that we can explore the modulation pattern only by changing the 
orientation of $\mathbf{q}$ with respect to the dimer axis.
Therefore, we studied two sample orientations. 
The first one uses a (110) surfaces with (110) and (001) lying in the scattering plane. 
The second sample features a (001) surface with (001) and (100) spanning the scattering plane. 
The sample orientation was determined by Laue diffraction and, for the sample with the (001)-(100) scattering plane, the additional observation of the (002) Bragg reflection. 

The numerical simulations of the excitation spectrum and of the RIXS intensities 
were performed using the {\sc Quanty} package \cite{Haverkort16}.

\noindent
\textbf{Acknowledgments}
\\
We dedicate this study to Daniel Khomskii, who has been a major source of inspiration 
for our work on cluster Mott insulators and beyond. 
We would like to acknowledge DESY -- a member of the Helmholtz Association HGF -- 
for access to beam time.
Furthermore, we acknowledge funding from the Deutsche Forschungsgemeinschaft 
(DFG, German Research Foundation) through Project No.\ 277146847 -- CRC 1238 
(projects A02, B03),  
Project No.\ 247310070 -- CRC 1143 (project A05), 
the W\"urzburg-Dresden Cluster of Excellence on Complexity and Topology in Quantum Matter -- ct.qmat (EXC 2147, project-id 390858490),
and by the Swedish Research Council through Project No.\ 2025-0409.

\appendix

\renewcommand{\theequation}{A\arabic{equation}}
\renewcommand{\thefigure}{A\arabic{figure}}

\setcounter{section}{0}
\setcounter{figure}{0}

\begin{figure}[t]
	\centering
	\includegraphics[width=0.98\columnwidth]{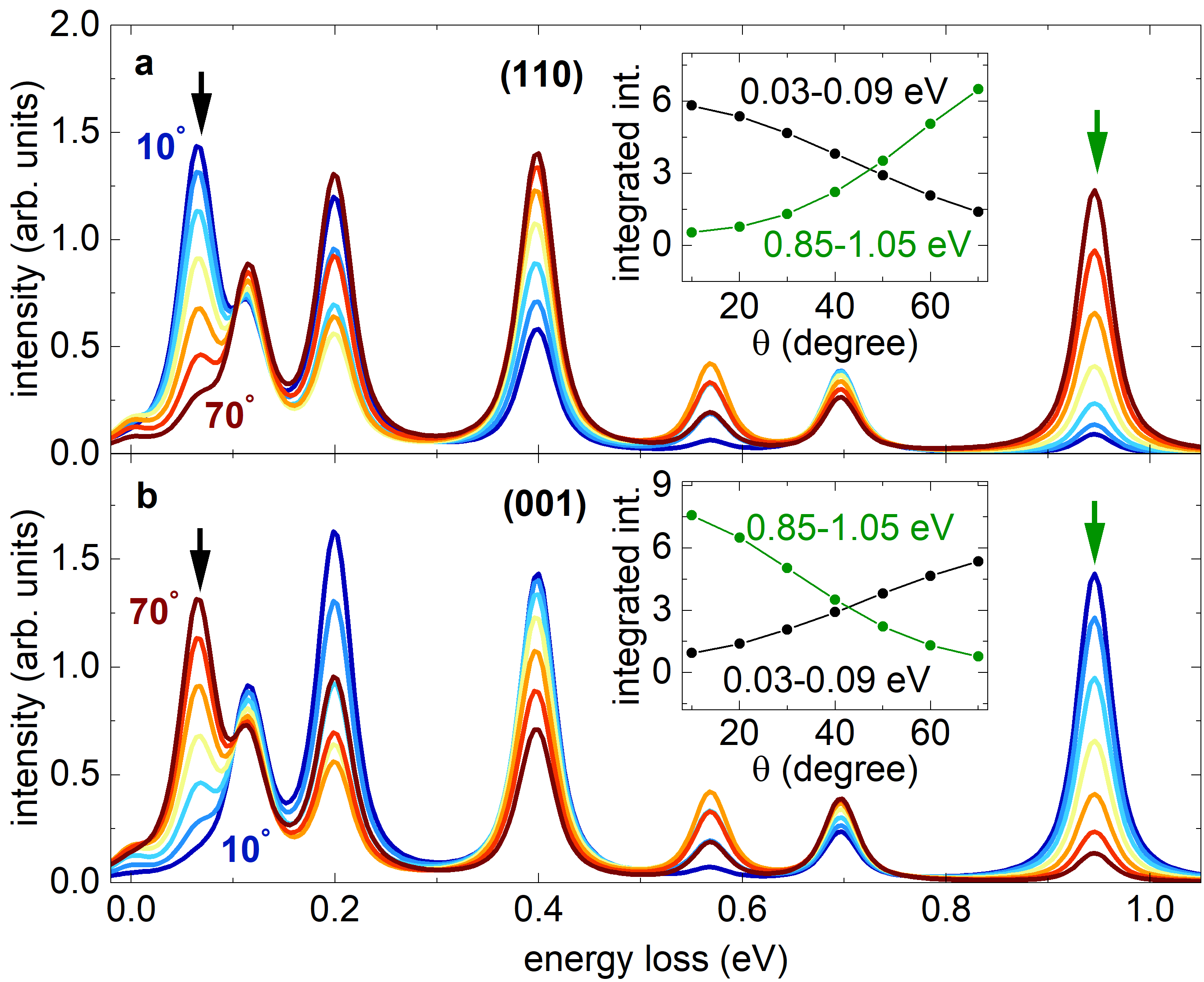}
	\caption{\textbf{Polarization dependence for a single site with two $t_{2g}$ holes.} 
    The two panels refer to the two distinct sample orientations, the (110) surface 
    and the (001) surface. With the exception of vanishing hopping, $t$\,=\,0, we use 
    the same parameters as for the dimers in the main text: 
    $U$\,=\,2\,eV, $J_{\rm H}$\,=\,0.26\,eV, $\Delta_{\rm trig}$\,=\,0.27\,eV, 
    $\zeta$\,=\,0.15\,eV.\@ 
    Insets: $\theta$ dependence of the RIXS intensity of the peaks marked by arrows.
}
	\label{fig:single_site_pol_dep}
\end{figure}

\section{RIXS calculations for a single site}
\label{sec:App_RIXS_single}

To address the polarization dependence, we calculated the RIXS intensity for a single $4d^4$ site 
with the same parameters that we have found for Ba$_3$CeRu$_2$O$_9$ with the exception of 
vanishing hopping. Furthermore, we consider the same scattering geometry and sample orientations 
as for the dimers. The results for a single site reproduce the main features of the polarization 
dependence that we observed for the dimers, see Fig.\ \ref{fig:single_site_pol_dep}. 
In particular, the intensity for the (110) orientation at 1\,eV is maximized for large $\theta$, 
while the lowest-energy peak shows the opposite trend, the intensity being maximized at small 
$\theta$. Moreover, the opposite behavior is observed for the (001) orientation. 
This strongly supports that the monotonic $\theta$ dependence of the RIXS intensity observed, 
e.g., around 0.5\,eV (see Fig.\ \ref{fig:integrals}) predominantly is caused by polarization. 
In contrast, the sinusoidal modulation as a function of $l$ is a fingerprint of the dimers.

\section{$j_z$ eigenstates for quantization along the dimer axis}\label{sec:jzEigenstates}
\label{sec:App_states}

\begin{figure}[t]
	\centering
	\includegraphics[width=.42\linewidth]{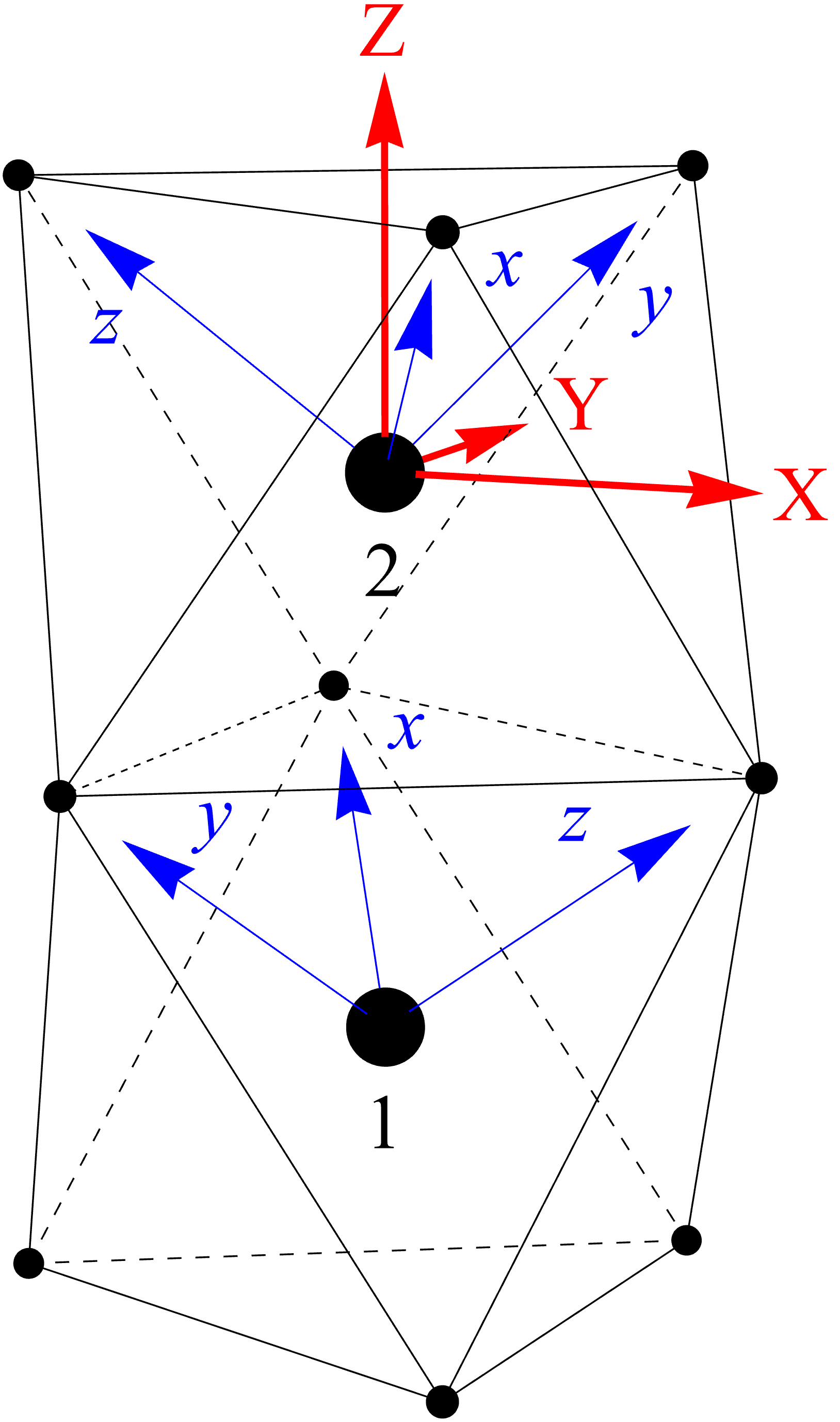}
	\caption{Sketch of local (blue) and global (red) coordinate frames.}
	\label{fig:dimer_frames}
\end{figure}

For the dimer, we distinguish between the global coordinate system and the two local ones, 
see Fig.\ \ref{fig:dimer_frames}.
The global or dimer frame is denoted using capital letters in the  subscripts, 
while $n$ indicates the  local frames for the lower 
octahedron (1) and upper  octahedron (2).
These frames are connected to each other by rotation matrices. 
A general vector $\mathbf{v}$  can be expressed in any of the three frames with
\begin{align}\label{eq:comp_vector}
\left(\begin{array}{c} v_X\\v_Y\\v_Z \end{array} \right) = R_1 \left(\begin{array}{c} v_x^{(1)}\\v_y^{(1)}\\v_z^{(1)} \end{array} \right) =  R_2 \left(\begin{array}{c} v_x^{(2)}\\v_y^{(2)}\\v_z^{(2)} \end{array} \right)
\end{align}
and the rotation matrices 
\begin{align}\label{eq:rotation_matrix}
	R_1=\left(
	\begin{array}{ccc}
		\frac{-1}{\sqrt{6}} & \frac{-1}{\sqrt{6}} & \sqrt{\frac{2}{3}} \\
		\frac{1}{\sqrt{2}} & \frac{-1}{\sqrt{2}} & 0 \\
		\frac{1}{\sqrt{3}} & \frac{1}{\sqrt{3}} & \frac{1}{\sqrt{3}} \\
	\end{array}
	\right)  
% \nonumber\\
    \, , \hspace{5mm}
	R_2=\left(
	\begin{array}{ccc}
		\frac{1}{\sqrt{6}} & \frac{1}{\sqrt{6}} & -\sqrt{\frac{2}{3}} \\
		\frac{-1}{\sqrt{2}} & \frac{1}{\sqrt{2}} & 0 \\
		\frac{1}{\sqrt{3}} & \frac{1}{\sqrt{3}} & \frac{1}{\sqrt{3}} \\
	\end{array}
	\right). 
\end{align}
Our aim here is to compute the $|j,j_Z\rangle$ eigenstates 
of the Hamiltonian
\begin{align}\label{eq:ham_J}
	\mathcal{H}&=\mathbf{L}\cdot \mathbf{S} 
\end{align}
for a single hole on the dimer with the \textit{dimer} axis 
as the quantization axis.
We use the general relations 
\eqref{eq:comp_vector}-\eqref{eq:rotation_matrix} 
to express the orbital operator $\mathbf{L}$ in terms of the local operators 
$L_x^{(n)}$, $L_y^{(n)}$, and $L_z^{(n)}$. 
To be explicit, we obtain for, e.g., the lower octahedron 1 
\begin{align}
	L_X &= -L_x^{(1)}/\sqrt{6} - L_y^{(1)}/\sqrt{6}+ \sqrt{2/3} L_z^{(1)}\nonumber\\
	L_Y &= L_x^{(1)}/\sqrt{2} - L_y^{(1)}/\sqrt{2}\nonumber\\
	L_Z&=(L_x^{(1)} + L_y^{(1)}+L_z^{(1)})/\sqrt{3} \,  .
\end{align}
We can now diagonalize \eqref{eq:ham_J} to obtain the explicit expressions 
of the $|j,j_Z\rangle$ eigenstates in terms of one of the standard bases 
for a single octahedron. It turns out that the ($a_{1g}$, $e^\pi_g$) basis 
\begin{align}
|a_{1g},\sigma\rangle_n = &\frac{1}{\sqrt{3}}\Big(|xy,\sigma\rangle_n +|yz,\sigma\rangle_n 
                           + |zx, \sigma\rangle_n \Big)\\
|e_{g\pm}^\pi,\sigma\rangle_n = & \pm\frac{1}{\sqrt{3}}\Big(|xy,\sigma\rangle_n \nonumber \\
         & +e^{\pm i \frac{2\pi}{3}}|yz,\sigma\rangle_n +e^{\mp i \frac{2\pi}{3}} |zx, \sigma\rangle_n \Big) \, , 
\end{align}
written in terms of the \textit{local} coordinate systems $n$\,=\,1,2, 
is the most convenient. 
Introducing the vectors 
\begin{align}\label{eq:basis_vectors}
\mathbf{u}_n&= \left( |a_{1g}\uparrow\rangle_n,|a_{1g}\downarrow\rangle_n, |e_{g+}^{\pi}\uparrow\rangle_n,\ldots,  |e_{g-}^{\pi}\downarrow\rangle_n \right)\nonumber\\
\mathbf{v}_n&= \left( \Big|\frac{1}{2},\frac{1}{2}\Big\rangle_n,\Big|\frac{1}{2},-\frac{1}{2}\Big\rangle_n \Big|\frac{3}{2},\frac{3}{2}\Big\rangle_n ,\ldots,  \Big|\frac{3}{2},-\frac{3}{2}\Big\rangle_n \right), 
\end{align}
we can express the $|j,j_Z\rangle$ eigenstates using
\begin{align}
\mathbf{v}_n^T &= J^{(n)} \mathbf{u}_n^T
\end{align}
with the transformation matrices 
\begin{align}  
% \label{eq:J^1_J^2}
    J^{(1)}&= \left(
\begin{array}{cccccc}
 -\frac{1}{\sqrt{3}} & 0 & 0 & -\sqrt{\frac{2}{3}} & 0 & 0 \\
 0 & \frac{1}{\sqrt{3}} & 0 & 0 & \sqrt{\frac{2}{3}} & 0 \\
 0 & 0 & -1 & 0 & 0 & 0 \\
 \sqrt{\frac{2}{3}} & 0 & 0 & -\frac{1}{\sqrt{3}} & 0 & 0 \\
 0 & \sqrt{\frac{2}{3}} & 0 & 0 & -\frac{1}{\sqrt{3}} & 0 \\
 0 & 0 & 0 & 0 & 0 & -1 \\
\end{array}
\right)\nonumber
\end{align}
\begin{align}  
\label{eq:J^1_J^2}
J^{(2)}&=\left(
\begin{array}{cccccc}
 -\frac{1}{\sqrt{3}} & 0 & 0 & \sqrt{\frac{2}{3}} & 0 & 0 \\
 0 & \frac{1}{\sqrt{3}} & 0 & 0 & -\sqrt{\frac{2}{3}} & 0 \\
 0 & 0 & 1 & 0 & 0 & 0 \\
 \sqrt{\frac{2}{3}} & 0 & 0 & \frac{1}{\sqrt{3}} & 0 & 0 \\
 0 & \sqrt{\frac{2}{3}} & 0 & 0 & \frac{1}{\sqrt{3}} & 0 \\
 0 & 0 & 0 & 0 & 0 & 1 \\
\end{array}
\right) \, .
\end{align}
In particular, 
we find $|3/2,3/2\rangle_n$\,=\,$(-1)^n |e_{g+}^\pi,\uparrow\rangle_n$ 
and $|3/2,-3/2\rangle_n$\,=\,$(-1)^n |e_{g-}^\pi,\downarrow\rangle_n$.
Note that the signs for the eigenstates in Eq.~\eqref{eq:J^1_J^2} are chosen such 
that $J^{\pm}$ has the standard form for both sites.

In terms of the  $j^2, j_Z$ eigenbasis, 
several terms in the dimer Hamiltonian look surprisingly  simple. 
We first consider hopping between octahedra 1 and 2, which is diagonal 
in the ($a_{1g}$,  $e_{g}^{\pi}$) basis: 
\begin{align}\label{eq:hopping}
\nonumber
	\mathcal{H}_{\text{hop}}&= -t_{a_{1g}} \Big(|a_{1g},\sigma\rangle_2 \,\, _1\langle a_{1g},\sigma| +f*|e_{g+}^\pi,\sigma\rangle_2 \,\, _1\langle e_{g+}^\pi,\sigma|\\
    &+ f*|e_{g-}^\pi,\sigma\rangle_2 \,\,  _1\langle e_{g-}^\pi,\sigma|\Big)+h.c. 
\end{align}
In the $j^2$, $j_Z$ basis, the same hopping Hamiltonian takes the form
\begin{widetext}
\begin{align}\label{eq:hopping_j}
    \mathcal{H}_{\rm hop} &= -\frac{t_{a_{1g}}}{3}\,\,    \mathbf{v}_1\, \left(
\begin{array}{cccccc}
 1-2 f & 0 & 0 & -\sqrt{2} (f+1) & 0 & 0 \\
 0 & 1-2 f & 0 & 0 & \sqrt{2} (f+1) & 0 \\
 0 & 0 & -3 f & 0 & 0 & 0 \\
 -\sqrt{2} (f+1) & 0 & 0 & 2-f & 0 & 0 \\
 0 & \sqrt{2} (f+1) & 0 & 0 & 2-f & 0 \\
 0 & 0 & 0 & 0 & 0 & -3 f \\
\end{array}
\right)  \mathbf{v}_2\,^\dagger + h.c. 
\end{align}
\end{widetext}
In particular, for the special case $f$\,=\,$-1$, the off-diagonal entries 
of \eqref{eq:hopping_j} vanish and hopping becomes diagonal with equal amplitudes. 
For arbitrary ratio $f$, hopping mixes the states $|j$\,=\,1/2,\,$j_Z$\,=\,$\pm 1/2\rangle$ 
with $|j$\,=\,3/2,\,$j_Z$\,=\,$\pm 1/2\rangle$.

The trigonal distortion on each octahedron can be captured by the Hamiltonian 
\begin{align}
	H_{\text{trig}}&= \frac{\Delta_{\text{trig}}}{3}\Big(-2|a_{1g},\sigma\rangle_n\, _n\langle a_{1g},\sigma| +|e_{g+}^\pi,\sigma\rangle_n \, _n\langle e_{g+}^\pi,\sigma|\nonumber\\
    &+|e_{g-}^\pi,\sigma\rangle_n \, _n\langle e_{g-}^\pi,\sigma| \Big). 
\end{align}
Rewriting it in terms of the $j$ eigenstates above, we find that trigonal distortions 
mix the states 
$|\frac{1}{2}, \pm \frac{1}{2}\rangle$ and $|\frac{3}{2}, \pm \frac{1}{2}\rangle$, 
while the states $|\frac{3}{2}, \pm \frac{3}{2}\rangle $ are only shifted in energy.
The corresponding Hamiltonian looks identical for both octahedra, $n$\,=\,1 and 2:
\begin{multline}
	H_{\text{trig}}= \frac{\Delta_{\rm trig}}{3}
    \mathbf{v}_n\,
	\left(
	\begin{array}{cccccc}
		0 & 0 & 0 & \sqrt{2} & 0 & 0 \\
		0 & 0 & 0 & 0 & -\sqrt{2} & 0 \\
		0 & 0 & 1 & 0 & 0 & 0 \\
		\sqrt{2} & 0 & 0 & -1 & 0 & 0 \\
		0 & -\sqrt{2} & 0 & 0 & -1 & 0 \\
		0 & 0 & 0 & 0 & 0 & 1 \\
	\end{array}
	\right)
	 \mathbf{v}_n^\dagger .
\end{multline}
Using the basis transformation \eqref{eq:J^1_J^2}, it is straightforward to see that 
\begin{multline}
  |a\!\uparrow\rangle_B \,
	|a\!\downarrow\rangle_B \,
	|-\!\uparrow\rangle_B \,
	|+\!\downarrow\rangle_B \\=    
    |\tfrac{1}{2},\tfrac{1}{2}\rangle_B \,
	|\tfrac{1}{2},-\tfrac{1}{2}\rangle_B \,
	|\tfrac{3}{2},\tfrac{1}{2}\rangle_B \,
	|\tfrac{3}{2},-\tfrac{1}{2}\rangle_B\, .
\end{multline}
For four holes per dimer, both correspond to Slater determinants where all 
single-particle states of the relevant subspace are occupied. 
Such determinants are invariant, up to an overall phase, under a unitary rotation 
of the underlying one-particle basis. 
Thus, $|\tilde\psi_0\rangle $ and consequently also $|\tilde \psi_4\rangle$ yield 
identical expressions whether or not the $(a_{1g},e_g^\pi)$ basis or the $j$ eigenbasis 
is used. This does not apply to $|\tilde{\psi}_2\rangle$. 
The symmetric linear combination containing two bonding and two antibonding orbitals 
does depend on the chosen basis.  
For the $j$ eigenbasis, the appropriate expression reads
\begin{align}
 |\tilde \psi_2\rangle &= \frac{1}{\sqrt{6}}\left(|\frac{1}{2},\frac{1}{2}\rangle_{AB} \,\,
   |\frac{1}{2},-\frac{1}{2}\rangle_{AB} \,\,
    |\frac{3}{2},\frac{1}{2}\rangle_B \,\,
   |\frac{3}{2},-\frac{1}{2}\rangle_B\right. \nonumber\\
   &\left.+|\frac{1}{2},\frac{1}{2}\rangle_{AB} \,\,
   |\frac{1}{2},-\frac{1}{2}\rangle_B \,\,
    |\frac{3}{2},\frac{1}{2}\rangle_{AB} \,\,
   |\frac{3}{2},-\frac{1}{2}\rangle_B +\ldots\right)\nonumber\\ 
\end{align}
and accounts for  11\,\% of the ED ground state, whereas Eq.\ \eqref{eq:bonding}, 
using the $(a_{1g},e_g^\pi)$ basis, carries  14\,\% instead.

\end{document}